\documentclass[onecolumn,preprint,superscriptaddress,nofootinbib,floatfix,longbibliography]{revtex4-1}
\usepackage[dvipsnames]{xcolor}
\usepackage{amssymb,amsmath,amsfonts,bm,slashed,soul,ragged2e,graphicx,hyperref,array,float}
\usepackage[utf8]{inputenc}
\usepackage{bm}
\usepackage[T1]{fontenc}
\usepackage[utf8]{inputenc}
\usepackage{amssymb}
\usepackage{amsfonts}
\usepackage{amsmath}
\usepackage{graphicx}
\usepackage{float}
\usepackage{relsize}
\usepackage{mathtools}
\usepackage{placeins}
\usepackage{array}
\usepackage{xcolor}
\usepackage{graphicx}%

\begin{document}

\title{Revisiting black holes surrounded by cloud and fluid of strings in general relativity}

\author{Luis C. N. Santos}
\email{luis.santos@ufsc.br}

\affiliation{Departamento de F\'isica, CFM - Universidade Federal de Santa Catarina; \\ C.P. 476, CEP 88.040-900, Florian\'opolis, SC, Brazil.}

\begin{abstract}
This paper revisits black hole solutions surrounded by clouds and fluids of strings within the framework of general relativity. We introduce a generalized equation of state for a fluid of strings with a variable parameter and derive a general solution to Einstein field equations for this system. We allow the parameter $\alpha$ in the equation of state for a fluid of strings ($\rho/p = \alpha$) to vary as a function of the radial coordinate. A particular solution with $M=0$ is explored, focusing on positive ranges of the equation of state parameter in analogy with the reduced Kiselev solution. Furthermore, we present a novel regular black hole solution that reduces to the Schwarzschild solution with a cloud of strings when the radial coordinate is much larger than a control parameter $r_0$. As an additional contribution of this work, we show that the energy-momentum tensor for the fluid of strings can be decomposed into contributions from three components: an isotropic perfect fluid, an electromagnetic field, and a scalar field minimally coupled. The paper examines the connection between the fluid of strings and Kiselev anisotropic fluid, revealing structural similarities through their energy-momentum tensors. The study also highlights how the obtained solutions  influence the horizons and thermodynamic properties of black holes. The new solutions allow for the emergence of geometries with multiple horizons and non-trivial temperature behaviors. As an additional test of the general solution, it is shown that particular choices for the function $\alpha(r)$ reproduce well-known results in the literature. Finally, we study geodesic motion of particles, geodesic completeness and shadows in the geometry of the novel regular black hole.

\end{abstract}


\maketitle

\preprint{}

\volumeyear{} \volumenumber{} \issuenumber{} \eid{identifier} \startpage{1} %
\endpage{}
\section{Introduction}

In recent years, the investigation of black holes surrounded by clouds and fluids of strings has gained considerable attention, enhancing our understanding of the interaction between gravitational fields and non-trivial matter distributions. These configurations provide significant extensions to classical black hole models by incorporating string-like matter, which can influence the black hole geometry, horizons, and thermodynamics. The foundational work by Letelier introduced the concept of a cloud of strings \cite{letelier1979clouds}, represented by a set of non-interacting one-dimensional objects that create a particular energy-momentum tensor. Later, developments extended this idea to fluids of strings by including pressure, yielding more complex solutions to Einstein field equations \cite{letelier1981fluids}. These models are especially relevant as they can allow new effects in the study, phenomena such as horizons, regularity, and thermodynamics of black holes. In this concern, several systems have been explored considering cloud of strings \cite{cloud1,cloud2,cloud3,cloud4,cloud5,cloud6,cloud7,cloud8,cloud11,cloud9,cloud10} and fluid of strings \cite{fluid1,fluid2,fluid3,fluid4,fluid5,fluid6,fluid7,fluid8}.
Spherically symmetric black holes represent a class of solutions to Einstein field equations characterized by their high degree of symmetry. The Schwarzschild metric with a cloud of strings, for example, describes a non-rotating, uncharged black hole with spherical symmetry. At the core of such solutions lies a singularity, which is an infinitesimal point where curvature becomes infinite. In this way, the inclusion of a surrounding fluid in Einstein field equations does not remove the singularity in most known solutions. 

Concerning the classical physics, it is possible to obtain a regular solution considering Einstein field equations and certain matter distributions in the space-time. Bardeen proposed a regular black hole model \cite{bardeen1968non} by replacing the mass
of Schwarzschild black hole by a function that depends on the radial coordinate.  Subsequent works expanded the original idea constructing regular space-times considering various theoretical approaches, such as, for example, Einstein field equations coupled with nonlinear electrodynamics \cite{ayon1998bardeen,bronnikov2001regular,han2020thermodynamics,bronnikov2020comment,junior2024regular,walia2024exploring,dolan2024superradiant,tangphati2024magnetically,guo2024recovery,kar2024novel,bronnikov2024regular}. 
Rotating regular black hole models can be developed starting from particular geometries, such as those of Hayward and Bardeen. The Newman–Janis algorithm provides a method to derive the space-time metric from these solutions \cite{bambi2013rotating}. More broadly, regular rotating black holes can also be formulated within the framework of nonlinear electrodynamics \cite{toshmatov2017generic}. In particular, it is possible to derive regular black holes as quantum corrections to classical singular black holes. These corrections may arise, for instance, from the influence of loop quantum gravity \cite{modesto2004disappearance,gambini2008black,ashtekar2023regular,modesto2006loop,momennia2022quasinormal,sharif2010quantum}.

The energy-momentum tensor of a fluid of strings for a particular space-time can effectively be interpreted as an anisotropic fluid, meaning that the pressure within the fluid varies along different spatial directions. This anisotropic behavior becomes relevant in the context of certain black hole solutions, such as the Kiselev black hole \cite{kiselev}, which generalizes the Schwarzschild solution by introducing a surrounding field modeled as a quintessential matter with an anisotropic energy-momentum tensor. In Kiselev black holes, the presence of anisotropy reflects the influence of a non-uniform pressure component, analogous to how a fluid of strings exhibits direction-dependent tension.  The anisotropic pressure plays an important role in Kiselev model, affecting the behavior of horizons and influencing the nature of the singularity, geodesics, and thermodynamics, as we can see from several studies in the literature \cite{shadow1,shadow2,shadow3,shadow4,shadow5,quasi1,quasi2,quasi3,quasi4,quasi5,quasi6,quasi7,quasi8,quasi9,quasi10,quasi11,termo1,termo2,termo3,termo4,termo5,termo6,termo7,termo8,termo9,termo10,termo11,termo12,termo13,termo14,termo15}. 

In this paper, we present four contributions to the study of fluids of strings and their connection to black hole solutions. First, we introduce an equation of state for a fluid of strings with a varying equation of state parameter and derive a general solution to Einstein field equations for this system.
Next, we examine a particular solution where the fluid of strings is simplified by setting \(M = 0\), focusing on a positive range for the equation of state parameter. Additionally, we obtain a novel solution for a regular black hole with the property that it reduces to the standard Schwarzschild solution in the presence of a cloud of strings when the radial coordinate \(r\) is much larger than a control parameter. 
Finally, we express the energy-momentum tensor for the fluid of strings as a combination of three types of energy-momentum tensors: those corresponding to a perfect fluid, an electromagnetic field, and a scalar field. 

The paper is organized as follows:
In Section \ref{sec2}, the concept of a generalized fluid of strings equation of state is introduced, along with its connection to the cloud of strings. This section explores the mathematical formulation of the energy-momentum tensor and develops solutions by varying the equation of state parameter. Section \ref{sec3} examines a formal connection between the fluid of strings and Kiselev anisotropic fluid. It shows how the two models are structurally linked through their energy-momentum tensors and investigates the physical implications of this connection, including changes in radial and tangential pressures. In Section \ref{sec4}, a new regular black hole solution is presented. This solution involves coupling a fluid of strings with an exponential term, ensuring regularity at the origin and recovering the Schwarzschild solution with cloud of strings at large radial distances. Section \ref{sec5} analyzes the decomposition of the energy-momentum tensor for the generalized fluid of strings. The tensor is expressed as a combination of perfect fluid, electromagnetic, and scalar field components. In section \ref{geodesic}, we analyze geodesic completeness and shadows of the regular black Hole. Finally, in Section \ref{sec6} are our conclusions. Throughout this work, we use units where $c = G = 1$.

\section{Generalized fluid of strings and Connection with cloud of String} \label{sec2}

In this section, we will briefly discuss the energy-momentum tensor related to a fluid of strings. In a first work,  Letelier \cite{letelier1979clouds} proposed a model to describe a cloud of strings, which is based on a surface bivector $\Sigma^{\mu\nu}$ that spans the two-dimensional time-like worldsheet of the strings. It can be defined as
\begin{equation}
    \Sigma^{\mu\nu}=\epsilon^{AB}\frac{\partial x^{\mu}}{\partial \xi^A}\frac{\partial x^{\nu}}{\partial \xi^B},
    \label{2.1}
\end{equation}
where $\epsilon^{AB}$ is the two-dimensional Levi-Civita symbol with components $\epsilon^{01}=-\epsilon^{10}=-1$. Note that the worldsheet has coordinates $\xi^A \in \left\{\xi^0,\xi^1 \right\}$, with $\xi^0$ and $\xi^1$ corresponding to time-like and space-like parameters, respectively.

A metric $h_{AB}$ induced on the worldsheet can be expressed in the form
\begin{equation}
   h_{AB}= g_{\mu\nu}\frac{\partial x^{\mu}}{\partial \xi^A}\frac{\partial x^{\nu}}{\partial \xi^B}.
    \label{eq2.2}
\end{equation}
Note that  $x(\xi^{A})$ describes the trajectory of the in strings world sheet. 
The energy-momentum tensor of a cloud of strings characterized by a proper density $\rho$ is given by \cite{letelier1979clouds}
\begin{equation}
T^{\mu\nu}=\rho\sqrt{-h}\frac{\Sigma^{\mu\lambda}\Sigma^{\;\nu}_{\lambda}}{(-h)},
 \label{2.4}
\end{equation}
where $h$ denote the determinant of the induced metric. Then, this model for a cloud of strings was later generalized by Letelier \cite{letelier1981fluids} to account for ``pressure''. In this case, the energy-momentum tensor is expressed as
\begin{equation}
    T^{\mu\nu}=(p+\rho\sqrt{-h})\frac{\Sigma^{\mu\lambda}\Sigma^{\;\nu}_{\lambda}}{(-h)}+pg^{\mu\nu},
     \label{2.5}
\end{equation}
where $p$ and $\rho$ are the pressure and density of the fluid of strings, respectively. 

Now, our goal is to solve Einstein field equations  coupled with a fluid of strings. To explicitly write the components of the energy-momentum tensor for the strings (\ref{2.5}), we need to consider a metric for the space-time, which will be solved together with the energy-momentum tensor. 
The metric describing a static, spherically symmetric space-time is expressed as:
\begin{equation}
    ds^2 = -f(r) dt^2 + \frac{1}{g(r)} dr^2 + r^2 \left( d\theta^2 + \sin{\theta}^2 d\varphi^2 \right).
    \label{2.6}
\end{equation}
For our system, the field equations imply that $f(r) = g(r),$ where the function $f(r)$ will be determined. As noted in \cite{soleng1995dark}, due to the symmetries of the metric, the only nonvanishing components of $\Sigma_{\mu\lambda}$ are $\Sigma_{tr}$ and $\Sigma_{\theta\varphi}$ and, furthermore, the induced metric $h$ satisfies $h < 0$. These conditions imply that the energy-momentum tensor for a fluid of strings reduces to
\begin{equation}
    T_{\: \: t}^{t} = T_{\: \: r}^{r}  \qquad \text{and} \qquad T_{\: \: \theta}^{\theta}=T_{\: \: \varphi}^{\varphi}= p.
    \label{2.7}
\end{equation}
To fully define the energy-momentum tensor, we must establish an equation of state for the fluid of strings. Here, we propose a generalization of the equation of state used in 
 \cite{soleng1995dark}, where energy density and pressure are related in the form $\rho(r) = \alpha p(r)$ with $\alpha$ being a constant. Instead, we consider that $\alpha$ depends on the radial coordinate, \textit{i.e,} $\alpha = \alpha(r)$. Thus, this generalized equation of state adds an additional difficulty to solving Einstein field equations, since there is now the presence of an indeterminate function $\alpha(r)$. But as we will see below, the existing symmetries of Einstein field equations and of the energy-momentum tensor allow us to obtain a general solution for a function $\alpha(r)$ where several known solutions can be reproduced as a special case (cloud of strings, fluid of strings with constant $\alpha$ and so on). Thus our equation of state for the fluid of strings can be written as   
 \begin{equation}
     \rho(r) = \alpha(r) p(r),
 \end{equation}
 and the associated energy-momentum tensor
\begin{equation}
    T^{\mu}_{\;\;\nu}=\left[- \rho(r),- \rho(r),\frac{\rho(r)}{\alpha(r)},\frac{\rho(r)}{\alpha(r)}\right].
    \label{2.8}
\end{equation}
In the case of the cloud of strings energy-momentum tensor, the angular pressures are zero. This can be achieved by taking the limit $\alpha \rightarrow \infty$ in Eq. (\ref{2.8}) so that $p \rightarrow 0$. By considering Einstein field equations  
\begin{equation}
    R^\nu_\mu - \frac{1}{2} \delta^\nu_\mu R = 8\pi T^\nu_\mu, 
    \label{2.9}
\end{equation}
 and the energy-momentum tensor (\ref{2.8}),  we find that the $(t,t)$ and $(r,r)$ components are identical and given by the following differential equation
\begin{equation}
    \frac{f'(r)}{r} + \frac{f(r)}{r^2} - \frac{1}{r^2} = -8\pi \rho(r).
    \label{3.10}
\end{equation}
Similarly, the $(\theta,\theta)$ and $(\varphi,\varphi)$ components are also identical and given by
\begin{equation}
    \frac{f''(r)}{2} + \frac{f'(r)}{r} = 8\pi \frac{\rho(r)}{\alpha(r)}.
     \label{3.11}
\end{equation}
By eliminating $\rho$ using both Eqs. (\ref{3.10}) and (\ref{3.11}), we arrive at the following equation:
\begin{equation}
     \frac{f'(r)}{r} + \frac{f(r)}{r^2} - \frac{1}{r^2}  = 
     - \alpha(r)\left( \frac{f''(r)}{2} + \frac{f'(r)}{r} \right).
     \label{3.12}
\end{equation}
Equation (\ref{3.12}) can be rewritten as follows:
\begin{equation}
    \frac{d}{dr} \left( r f(r) \right) - 1 = \frac{a(r) r}{2}  \frac{d}{dr} \left( \frac{d}{dr} \left( r  f(r) \right) - 1 \right)
    \label{3.13}
\end{equation}
which can be easily integrated to obtain:
\begin{equation}
     \frac{d}{dr} \left( r f(r) \right) - 1  = c_1 e^{\left( \int \left( \frac{-2}{a(r) r} \right) dr \right)}.
    \label{3.14}
\end{equation}
As a particular solution, the case $\alpha = 2$ must be analyzed separately. In the general solution that we will obtain for the function $f(r)$, this particular value for $\alpha$ is associated with a singularity. In the case of Eq. (\ref{3.14}) the value $\alpha = 2$ lead to a logarithmic solution in the form 
\begin{equation}
    f(r) = 1 - \frac{2c}{r} + \frac{d\ln{\lambda r}}{r},
    \label{3.15}
\end{equation}
where $c,d$ and $\lambda$ are constants. In this way, we have two classes of solutions, $\alpha = 2$ and $\alpha \neq 2$. This type of logarithmic solution is important for modeling rotation curves in galaxies since they can lead to a constant rotation curve at large distances from the center. 

Now considering a general solution, \textit{i.e.}, $\alpha \neq 2$, and integrating once again Eq. (\ref{3.14}), we obtain the general solution for the strings fluid with variable equation of state:
\begin{equation}
    f(r) = 1 + \frac{c_2}{r} + \frac{c_1  \int e^{\left( \int \left( \frac{-2}{a(r) r} \right) dr \right)} dr}{r},
    \label{3.16}
\end{equation}
where $c_2 = -2M$ and $c_1$ an arbitrary integration constant. In the absence of fluid, this equation reduces to the Schwarzschild solution. For the particular case where $\alpha$ is a constant, the integrals can be easily solved providing the solution 
\begin{equation}
    f(r) = 1 - \frac{2M}{r} + \frac{c_1 \alpha r^{\frac{-2}{\alpha}}}{\alpha -2},
    \label{3.17}
\end{equation}
corresponding the solution obtained in \cite{soleng1995dark} with the usual equation of state for $\alpha \neq 2$. If $\alpha \rightarrow \infty$, Eq. (\ref{3.16}) can be solved in the form
\begin{equation}
    f(r) = 1 - \frac{2M}{r} - \epsilon,
    \label{3.18}
\end{equation}
that is the solution for a cloud of strings \cite{letelier1979clouds}, where $\epsilon$ is a constant. Thus, by choosing the function $\alpha(r)$ appropriately, we are able to obtain known solutions associated with different types of physical systems.
Furthermore, we can use this approach to generate new solutions for Einstein field equations  by choosing the function of the equation of state appropriately. It is worth noting that an equation of state with a variable parameter was used in the context of the Kiselev fluid \cite{santos2024regular}, where several classes of regular black holes were related to this parameter and in the context of wormholes in $f(R,T)$ gravity \cite{rastgoo2025wormholes}. 

\section{Connecting fluid of strings and Kiselev anisotropic fluid} \label{sec3}

A class of solution of Einstein field equations  corresponding to a black hole surrounded by a type of anisotropic effective fluid was obtained by Kiselev \cite{kiselev}. In this system, the energy-momentum tensor is defined to have the components of the spatial sector proportional to the time sector \cite{santos2023kiselev,da2024black} in the form
\begin{align}
    T^{t}_{\:\:\:t}= T^{r}_{\:\:\:r}&=- \rho(r), \label{e8}\\
   T^{\theta}_{\:\:\:\theta}= T^{\phi}_{\:\:\:\phi}&= \frac{1}{2}\rho(r) (3w+1),
   \label{e9}
\end{align}
where $w$ is a constant parameter of the equation of state of the fluid.
 Equations (\ref{e8}) and (\ref{e9}) 
 have the components of  energy-momentum tensor effectively connected to an anisotropic fluid  represented by 
\begin{equation}
    T^{\mu}_{\:\:\:\nu} = diag(-\rho,p_r,p_t,p_t),
    \label{e10}
\end{equation}
where $p_r=-\rho$ is the radial pressure and $p_t=\frac{1}{2}\rho (3w+1)$ is the tangential pressure. By using these components of Kiselev anisotropic fluid into Einstein field equations, we obtain the solution for a Kiselev black hole
\begin{equation}
   f(r) =  1 - \frac{2M}{r} - \sum_n \left( \frac{r_n}{r} \right)^{3w_n + 1}
\end{equation}
where $r_n$ is a constant, $n \in\mathbb{N}^*$ and $w_n \neq \{0, \frac{1}{3}, -1\}$ for a particular value of $n$. Comparing Eqs. (\ref{e8}) and (\ref{e9}) with (\ref{2.8}), we obtain a correspondence between the Kiselev energy-momentum tensor and the strings fluid by identification: $\alpha \rightarrow 2/(3w+1)$.
\subsection{Reduced Kiselev/strings black hole: $-\frac{1}{3} < w < 0$}
\begin{figure*}[t]
    \centering    \includegraphics[width=0.45\textwidth]{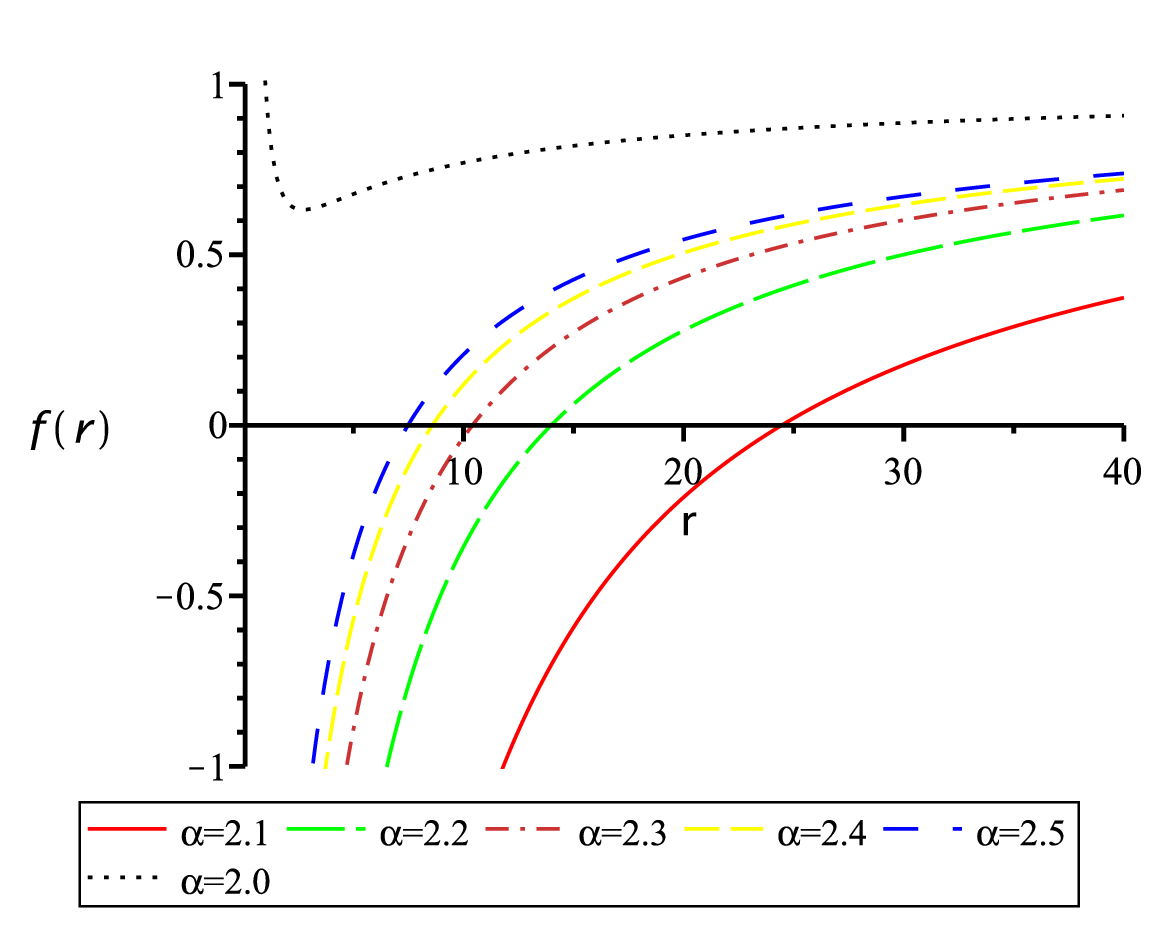}    \includegraphics[width=0.45\textwidth]{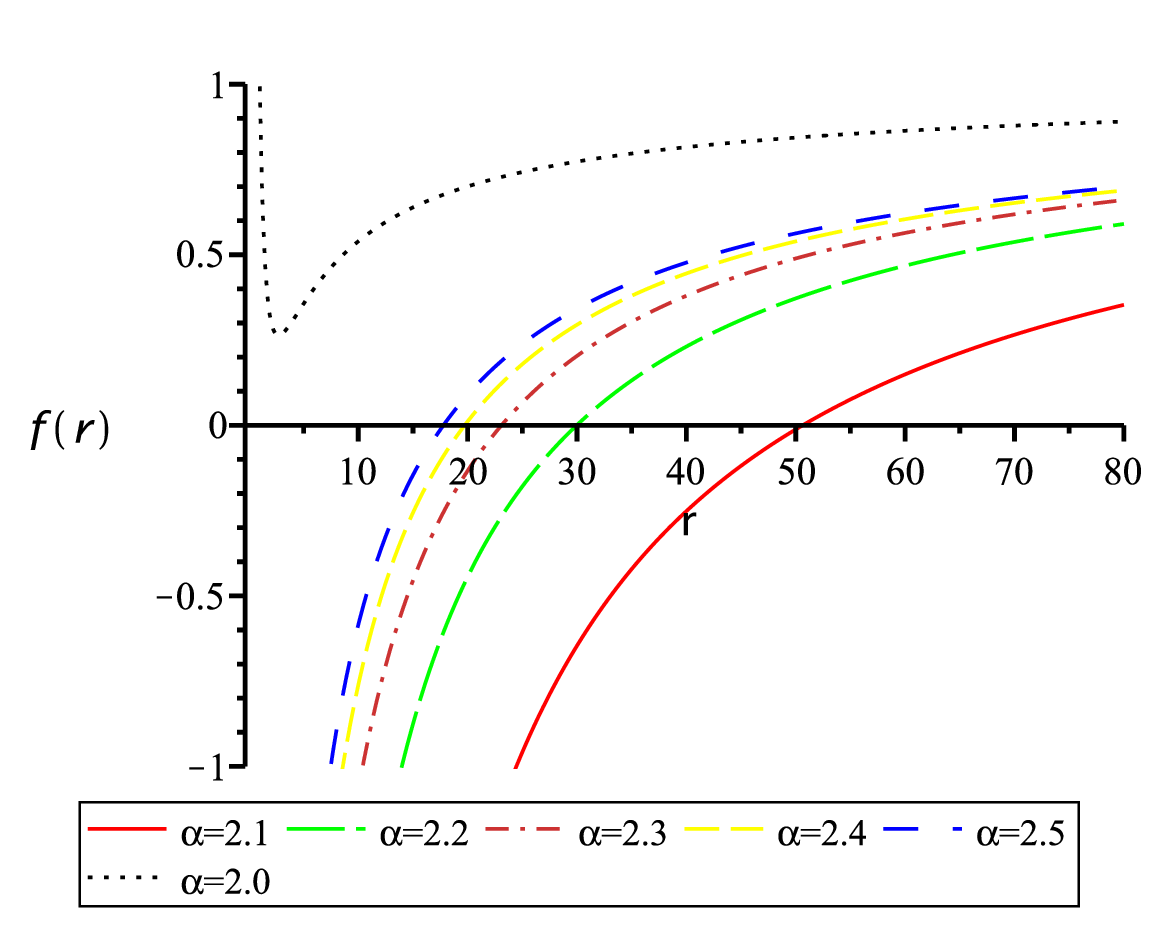}\\
\vspace{1cm}    \includegraphics[width=0.45\textwidth]{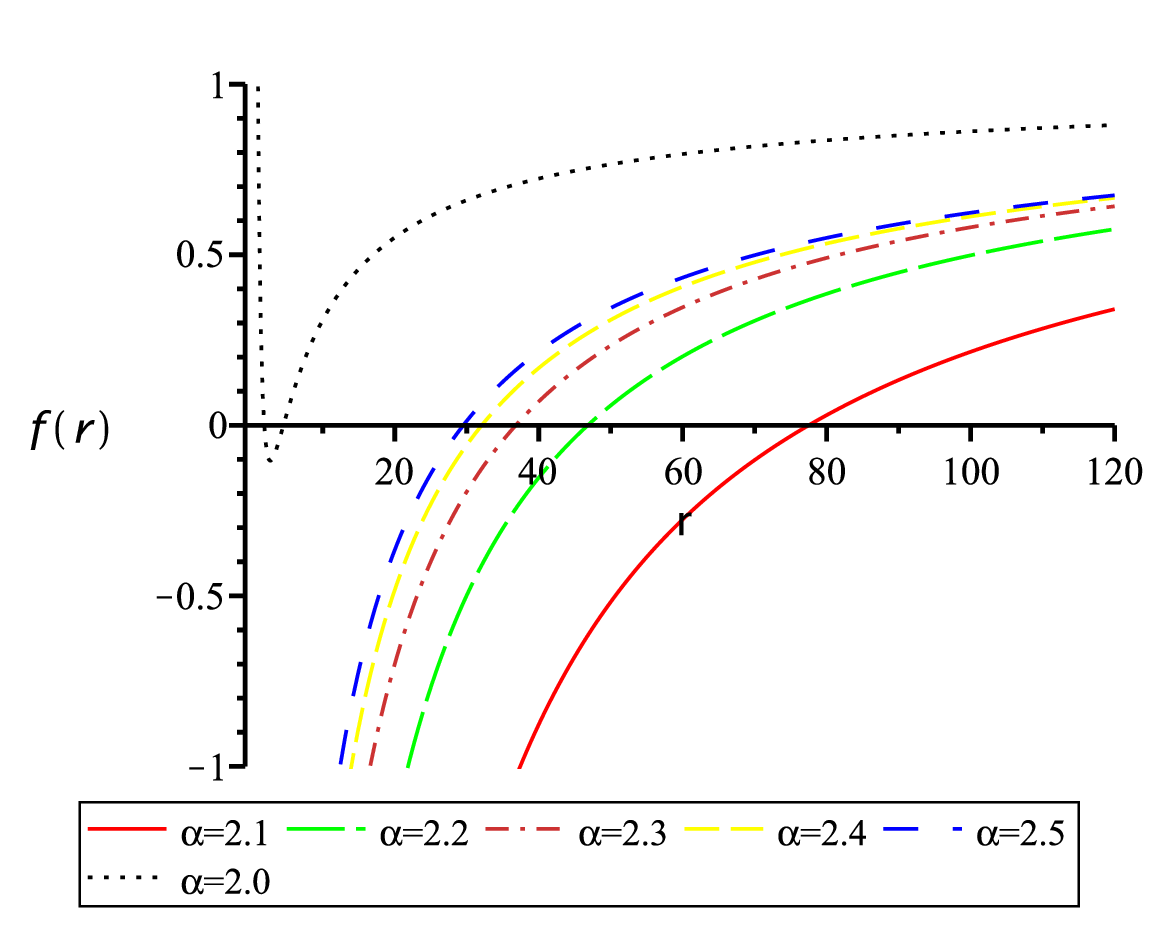}  \includegraphics[width=0.45\textwidth]{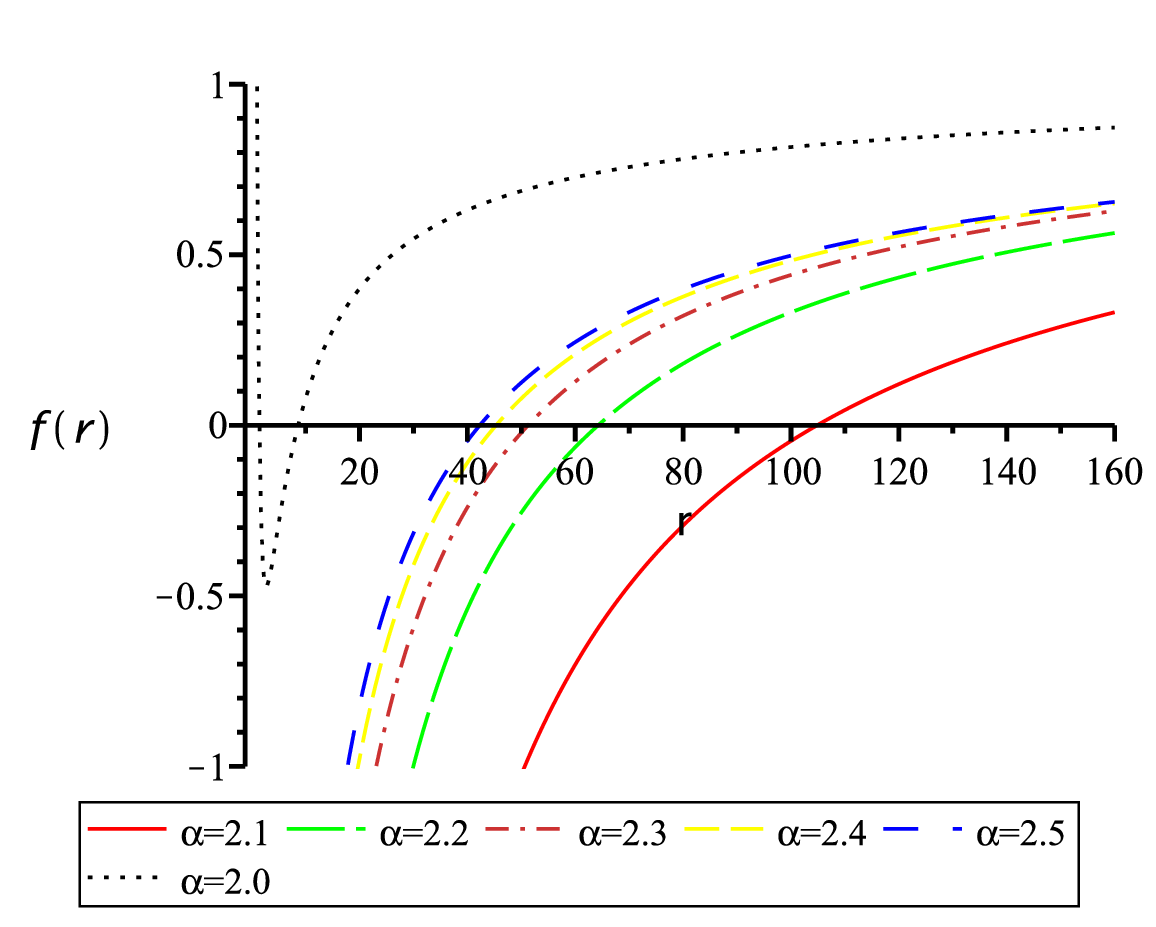}
    \caption{The figure presents four plots of the function \( f(r) \) as a function of \( r \), with curves corresponding to different values of the parameter \( \alpha \), ranging from 2.0 to 2.5 for \(l = \lambda =1\). Each plot in the figure corresponds to a different value of the parameter \( \epsilon \). The plot in the top-left position has \( \epsilon = 1 \), the top-right plot has \( \epsilon = 2 \), the bottom-left plot has \( \epsilon = 3 \), and the bottom-right plot has \( \epsilon = 4 \). As \( \epsilon \) increases, there are noticeable changes in the behavior of \( f(r) \) and the corresponding positions of the horizons, with these variations becoming more evident across the different values of \( \epsilon \) in each plot. The curves are distinguished by color and line style, as indicated in the legend below each plot: the black dotted line corresponds to \( \alpha = 2.0 \), the red solid line to \( \alpha = 2.1 \), the orange dashed line to \( \alpha = 2.2 \), the yellow dash-dotted line to \( \alpha = 2.3 \), the green dash-dot-dotted line to \( \alpha = 2.4 \), and the blue long-dashed line to \( \alpha = 2.5 \). The curves show an overall increasing behavior of \( f(r) \) as \( r \) grows, with more noticeable differences between the curves for different values of \( \alpha \) particularly for \( r > 10 \). In each plot, considering that horizon is defined as the value of \( r \) where the curve crosses the line \( f(r) = 0 \), the position of the horizon shifts as \( \alpha \) increases, with smaller values of \( \alpha \) resulting in larger horizons (non-logarithmic solutions). For \( \alpha = 2.0 \) (the black dotted curve, logarithmic solution), the horizon occurs at a smaller value of \( r \). }
    \label{f1}
\end{figure*}
\begin{figure}[t]
    \centering    
    \includegraphics[scale=0.4]{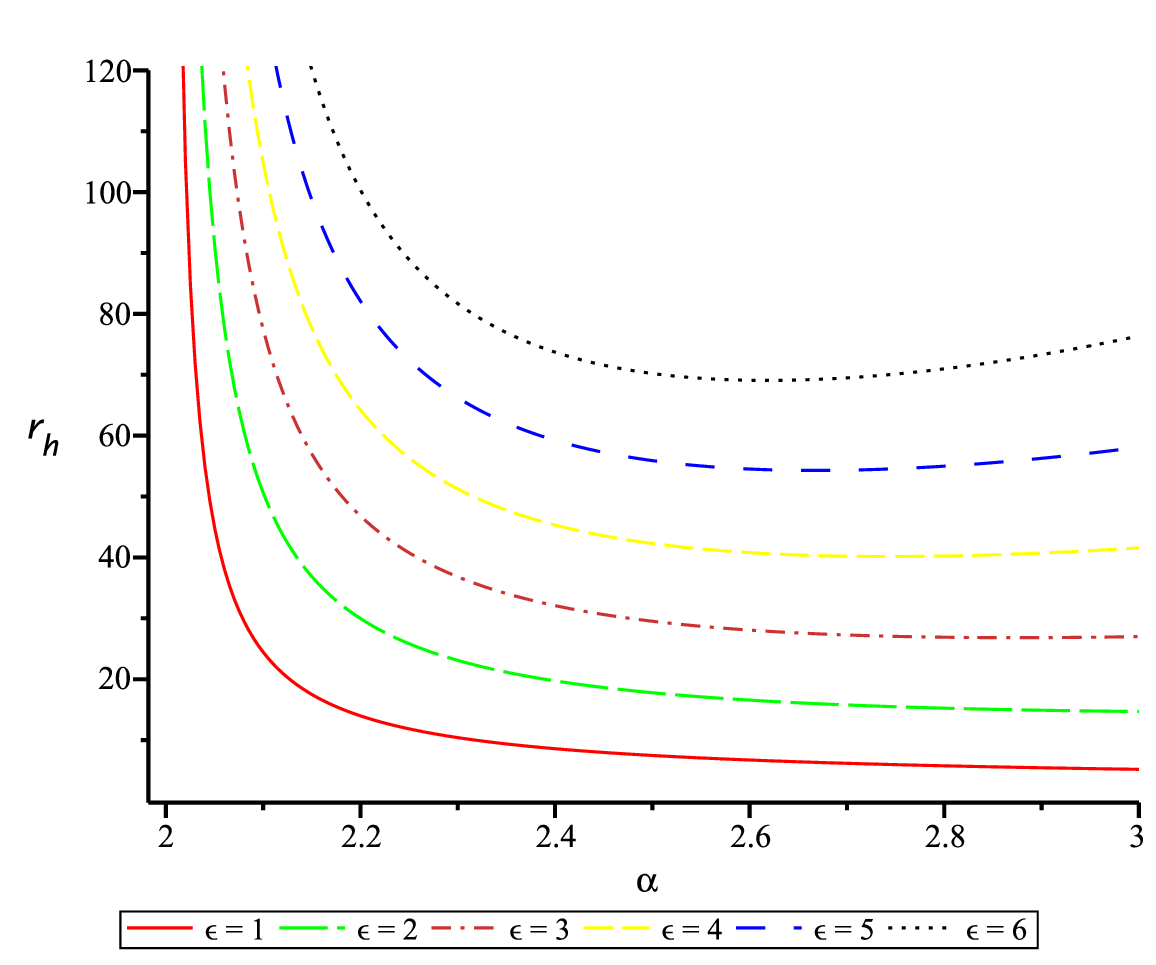}  \caption{Plot of the horizon $r_h$ as a function of $\alpha$. Considering values in the range $2 < \alpha < \infty$ with \(l = 1\), we study the behavior of the fluid of strings. In this plot, we used four values for the parameter $\epsilon$ associated with the energy density. The black dotted line corresponds to \( \epsilon = 1 \), the blue dashed line to \( \epsilon = 2 \), the yellow dash-dotted line to \( \epsilon = 3 \), the red dash-double-dotted line to \( \epsilon = 4 \), and the green solid line to \( \epsilon = 5 \). As we can see, the values of energy density where \(\alpha \rightarrow 2\) are associated with large values for the horizon radius.}
    \label{f2}
\end{figure}

\begin{figure}[t]
    \centering    
    \includegraphics[scale=0.4]{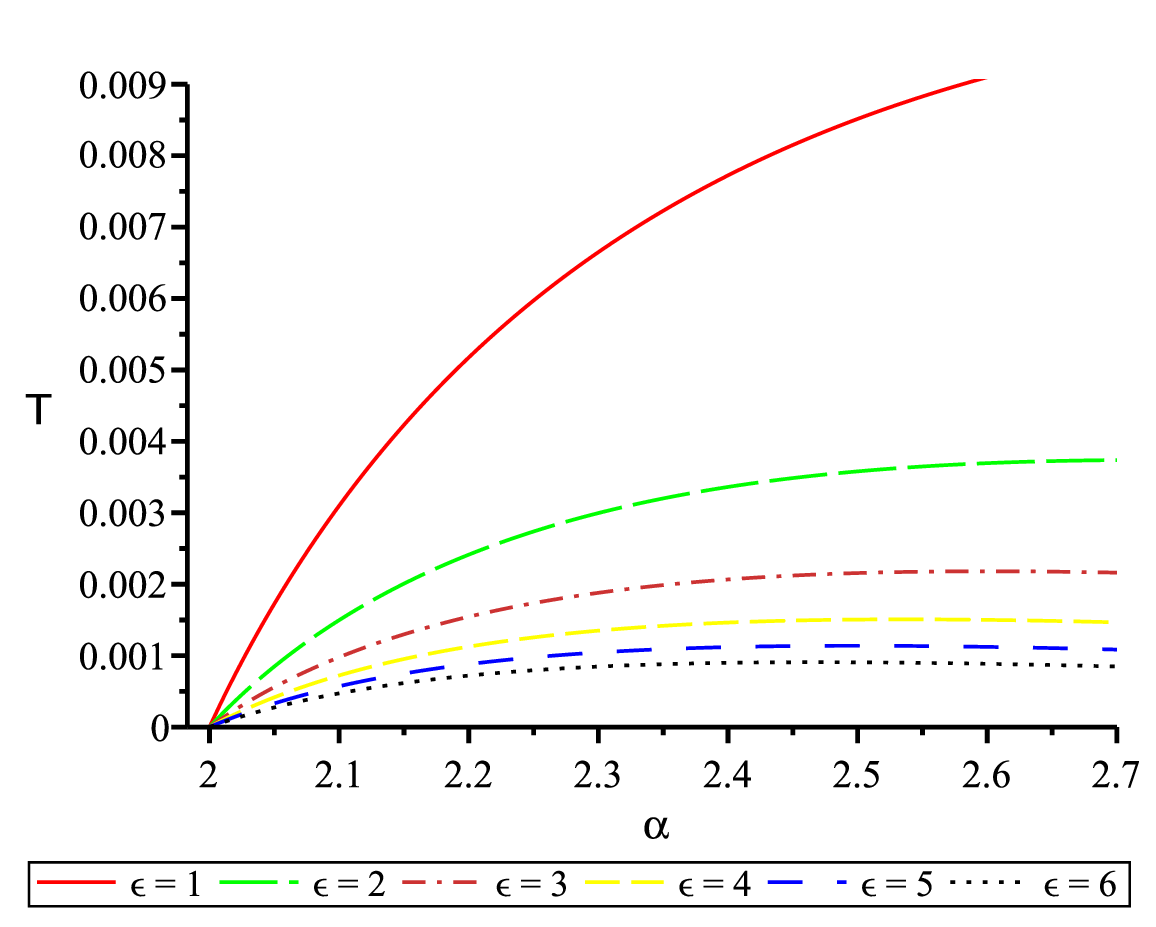}  \caption{The graph depicts the relation between Hawking Temperature (T) and the state parameter \(\alpha\). Several curves are shown, each corresponding to different values of \(\epsilon\) (ranging from 1 to 6, as indicated in the legend). As \(\alpha\) increases from 2 to approximately 2.7, the temperature T increases non-linearly for all values of \(\epsilon\), with the rate of increase depending on the value of \(\epsilon\). The curve for \(\epsilon = 1\) (red solid line) shows the steepest rise, while curves for larger \(\epsilon\) (such as \(\epsilon = 6\), represented by the black dotted line) demonstrate a slower growth in temperature. The general trend indicates that for higher values of \(\epsilon\), the Hawking Temperature is lower for a given value of \(\alpha\), with the temperature rising more gradually compared to smaller \(\epsilon\) values. We use \(l = \hbar =1\).}
    \label{f3}
\end{figure}

As a particular case, we analyze the black hole with fluid of strings associated with the reduced Kiselev black hole, defined by conditions $2M = 0 \ \text{and} \ -\frac{1}{3} < w < 0$ \cite{qu2023reduced}. As a result of these conditions, a new horizon type emerge and the tangential pressure changes sign. In the fluid of strings, the tangential pressure changes sign in a different region. 
We consider that the parameter of the equation of state of the fluid of strings assumes values in the range $0 < \alpha < \infty$. The solution of fluid of strings associated with this range and with $2M = 0$ can be referred as reduced fluid of strings solution.
As discussed before, the case $\alpha \rightarrow \infty$ is associated with the cloud of strings and the case $\alpha \rightarrow 2$ corresponds to the particular logarithmic solution. To make the solution associated with $\alpha$ constant more transparent, we can write Eq. (\ref{3.17}) and the logarithmic solution as (\cite{soleng1995dark}) 
\begin{eqnarray} \label{eq3a17}
f(r) =1 - \left\{
\begin{array}{cc}
\epsilon l r^{-1} \log(\lambda r) & \mbox{ for} \quad \alpha=2, \\ 

\epsilon \alpha( \alpha-2)^{-1} \left( \frac{l}{r}\right)^{2/\alpha} & \mbox{ for} \quad \alpha \neq 2,
\end{array}
\right.
\end{eqnarray}
where $l$ and $\lambda$ are positive constants, $\epsilon = \pm 1$ gives the sign of the energy density. Let us analyze the horizons of this space-time. In Fig. \ref{f1}, we show the solution $f(r)$ as a function of $r$. One can see that large values of $\epsilon$ increase the horizon radius. In contrast, large values of $\alpha$ decrease the horizon radius (non-logarithmic solutions). Thus, by considering the function for $\alpha \neq  2$, the condition $f(r_h) = 0$ give us the horizons associated with this metric, where $r_h$ is the horizon radius. The result is given by
\begin{equation}
    r_h = l \left( \frac{\alpha - 2}{\alpha \epsilon} \right)^{-\frac{\alpha}{2}}.
\end{equation}
So, in the case of a reduced fluid of strings solution, we can obtain an explicit expression for the horizons of the space-time. In Fig. \ref{f2}, we show the solution $f(r)$ as a function of $r_h$ for a fixed value of $l$. The plot shows the relation between the horizon radius \( r_h \) and the parameter \( \alpha \) for various values of the parameter \( \epsilon \), represented by different line styles and colors.  As \( \alpha \) increases, the horizon radius \( r_h \) decreases for some values of \( \epsilon \), exhibiting an asymptotic behavior as \( \alpha \) approaches higher values. The dependence of \( r_h \) on both \( \alpha \) and \( \epsilon \) suggests a consistent trend in which \(\alpha \rightarrow 2\) results in higher horizon radii, but the overall shape of the curves maintain a similar pattern for each \( \epsilon \). The surface gravity $\kappa=\left.\frac{1}{2}\frac{df(r)}{dr}\right|_{r=r_h}$ of the function \(f(r)\) associated to the temperature $T$ by the expression $T=\hbar\kappa/2\pi$, results in the following equation
\begin{equation}
    T = \frac{\hbar \epsilon}{2\pi l (\alpha - 2) } \left( \frac{\alpha - 2}{\alpha \epsilon} \right)^{\left( \frac{\alpha}{2} + 1 \right)},\:\:\alpha \neq 2
\end{equation}
and 
\begin{equation}
   T = \frac{\hbar \epsilon l}{4 r^2 \pi} \left( \ln(\lambda r) - 1 \right),\:\:\alpha = 2
\end{equation}
 Figure \ref{f3}, illustrates the connection between Hawking Temperature $T$ and the equation of state parameter \(\alpha\). The curves correspond to different values of \(\epsilon\) (ranging from 1 to 6, as specified in the legend). As \(\alpha\) increases from 2 to approximately 2.7, the temperature T increases non-linearly for all \(\epsilon\) values, with the growth rate depending on \(\epsilon\). The curve for \(\epsilon = 1\) (solid red line) shows the most rapid increase, while higher \(\epsilon\) values (e.g., \(\epsilon = 6\), represented by the black dotted line) demonstrate a slower rise. In general, larger \(\epsilon\) values correspond to lower temperatures for a given \(\alpha\), with a more gradual increase compared to smaller \(\epsilon\) values. 

\begin{figure}[t]
    \centering    
    \includegraphics[scale=0.4]{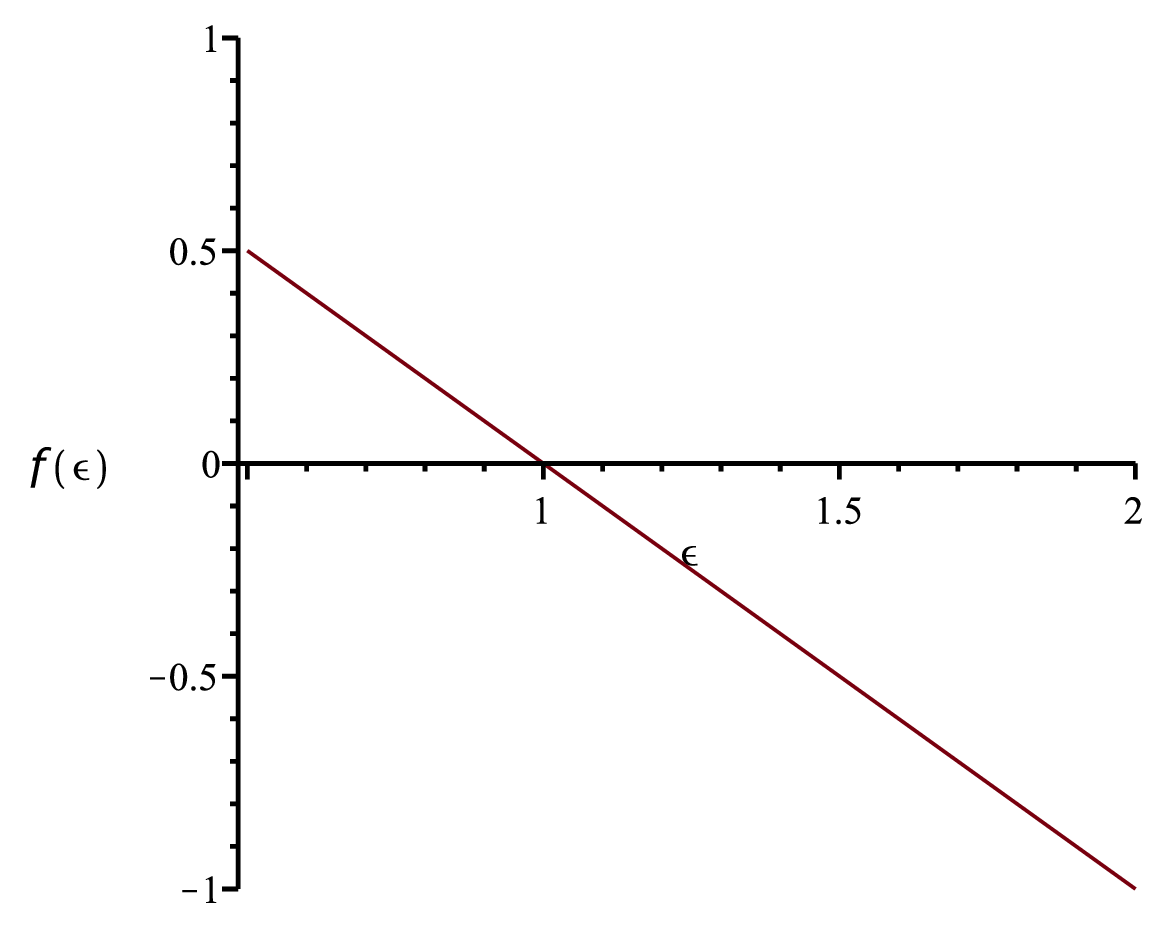}  \caption{The graph shows a solution related to a cloud of strings , plotting \( f(\epsilon) \) as a function of \( \epsilon \). The curve decreases linearly, and the point where the graph crosses the \(\epsilon\)-axis occurs at \( \epsilon = 1 \). The solution depicted by this graph does not depend on the radial coordinate \( r \), emphasizing that the relation between \( f(\epsilon) \) and \( \epsilon \) is independent of \( r \).}
    \label{f4}
\end{figure}

In the case of the reduced black holes with cloud of strings, we can obtain the solution describing this system by considering the limit \(\alpha \rightarrow \infty\). Thus, the function \(f(r)\) associated with the reduced solution can be written as
\begin{equation}
  f(\epsilon) = 1 - \epsilon.  
\end{equation}
This function crosses the $\epsilon$-axis at point $\epsilon =1$. In this situation, the space-time does not have horizons, but it has naked singularity at $r=0$ \cite{letelier1979clouds}.   Figure \ref{f4} illustrates a solution associated with a strings cloud, where \( f(\epsilon) \) is plotted against \( \epsilon \). The function decreases in a straight line, intersecting the \( \epsilon \)-axis at \( \epsilon = 1 \). This solution is independent of the radial coordinate \( r \), indicating that the relation between \( f(\epsilon) \) and \( \epsilon \) remains unchanged by variations in \( r \).

\section{New regular solution} \label{sec4}

The general form of Eq. (\ref{3.16}) suggests the search for expressions for the function $\alpha(r)$ in order to obtain new solutions of Einstein field equations. In what follows, we will investigate some forms for the equation of state that are associated with solutions of Einstein equation with certain physical properties. Initially, we will propose a new solution of Einstein equation corresponding a regular black hole surrounded by a fluid of strings, which can be associated with an effective  geometry of usual singular black hole (at \(r \rightarrow 0\)) surrounded by strings in the limit of large $r$. We will show that this solution mimics a regular cloud of strings. 
\subsection{Regular black hole surrounded by fluid}
Regular black holes can be constructed solving the classical gravitational field equations by incorporating an energy-momentum tensor related to a specific matter distribution. In this sense, Ricci curvature $R = g^{\mu\nu}R_{\mu\nu}$, the contraction of two Ricci tensors $R_{\mu\nu}R^{\mu\nu}$, and the
Kretschmann scalar $K =R_{\mu\nu\alpha\beta}R^{\mu\nu\alpha\beta}$ can be used as a decent test for regular solutions. Let us consider the formalism developed so far. The first step is to choose a particular form of the function $f(r)$. Then, the varying equation of state function must be chosen so that the solution of the integral in (\ref{3.16}) yields the desired form of the solution. Considering a interesting class of regular black holes obtained by Dymnikova \cite{dymnikova1992vacuum}, where the field equations are solved in the presence of a particular anisotropic fluid. This solution can be written as
\begin{equation}
    B(r) = 1 - \frac{r_g}{r}\left(1-\exp\left(-\frac{r^3}{r_0^3}\right) \right), 
    \label{301}
\end{equation}
where $r_0$ is a positive constant and $r_g = 2M$ with $M$ being the black hole mass. This geometry generate a black hole solution regular at $r =0 $ and everywhere else. For $r \gg r_0$, the solution coincides with the Schwarzschild solution. 

Naively, we can attempt to construct a solution associated with a regular black hole surrounded by a cloud of strings simply by adding the constant term $\epsilon$ that is associated with the solution for cloud of strings into Eq. (\ref{301}). However, the black hole resulting from this assumption is not regular, as can be seen by calculating the curvature invariants. To achieve our goal, we can couple the constant $\epsilon$ associated with the cloud of strings, with an exponential term similar to Eq. (\ref{301}), resulting in the following geometry
\begin{align}
    f(r) =& 1 - \frac{r_g}{r} \left(1-\exp\left(-\frac{r^3}{r_0^3}\right) \right)\nonumber\\
    &- \epsilon\left(1-\exp\left(-\frac{r^3}
    {r_0^3}\right) \right)\nonumber\\
    =& 1 - \left(\frac{r_g}{r} + \epsilon\right) \left(1-\exp\left(-\frac{r^3}{r_0^3}\right) \right).
    \label{302}
\end{align}
Equation (\ref{302}) describes a black hole with an additional term associated with the anisotropic fluid.  For $r \gg r_0$, this solution reduces to the singular Schwarzschild black hole with cloud of strings:  
\begin{equation}
    f(r) = 1 - \frac{2M}{r} - \epsilon.
    \label{303}
\end{equation}
The curvature invariants for Eq. (\ref{302}) in the limit where $r \rightarrow 0$, are
\begin{equation}
    R=-\frac{12r_g}{r_{0}^3},\:\:R_{\mu\nu}R^{\mu\nu} =\frac{36r_g^2}{r_{0}^6},\:\:K =\frac{24r_g^2}{r_{0}^6},
\end{equation}
stating the regularity of the space-time. To obtain solution  (\ref{302}) from  Einstein field equation, we can consider the following equation of state function
\begin{equation}
\alpha(r) = \frac{2 r_0^3 \left( \frac{\epsilon r_0^3 \exp\left(\frac{r^3}{r_0^3}\right)}{3} + \epsilon r^3 - \frac{\epsilon r_0^3}{3} + r_g r^2 \right)}{3 r^2 \left( \epsilon r^4 - \frac{4}{3} \epsilon r r_0^3 + r^3 r_g - \frac{2}{3} r_0^3 r_g \right)},
\label{extra1}
\end{equation} 
substituting expression (\ref{extra1}) into Eq. (\ref{3.16}), the integral can be calculated exactly and by choosing the constants of integration as $c_1 = -1/r_0^3$ and $c_2 =-r_g$, we arrive at eq. (\ref{302}). The energy density associated with this equation of state is derived from Einstein field equations. Then, substituting the solution (\ref{3.16}) and the function $\alpha(r)$ into (\ref{3.10}) or (\ref{3.11}), one finds
\begin{equation}
    \rho(r) = \frac{ \left( \epsilon r_0^3 \exp\left(\frac{r^3}{r_0^3}\right) + 3\epsilon r^3 - \epsilon r_0^3 + 3 r_g r^2 \right)}{8\pi r_0^3 \exp\left(\frac{r^3}{r_0^3}\right) r^2},
\end{equation}

\begin{figure*}[t]
    \centering    \includegraphics[width=0.45\textwidth]{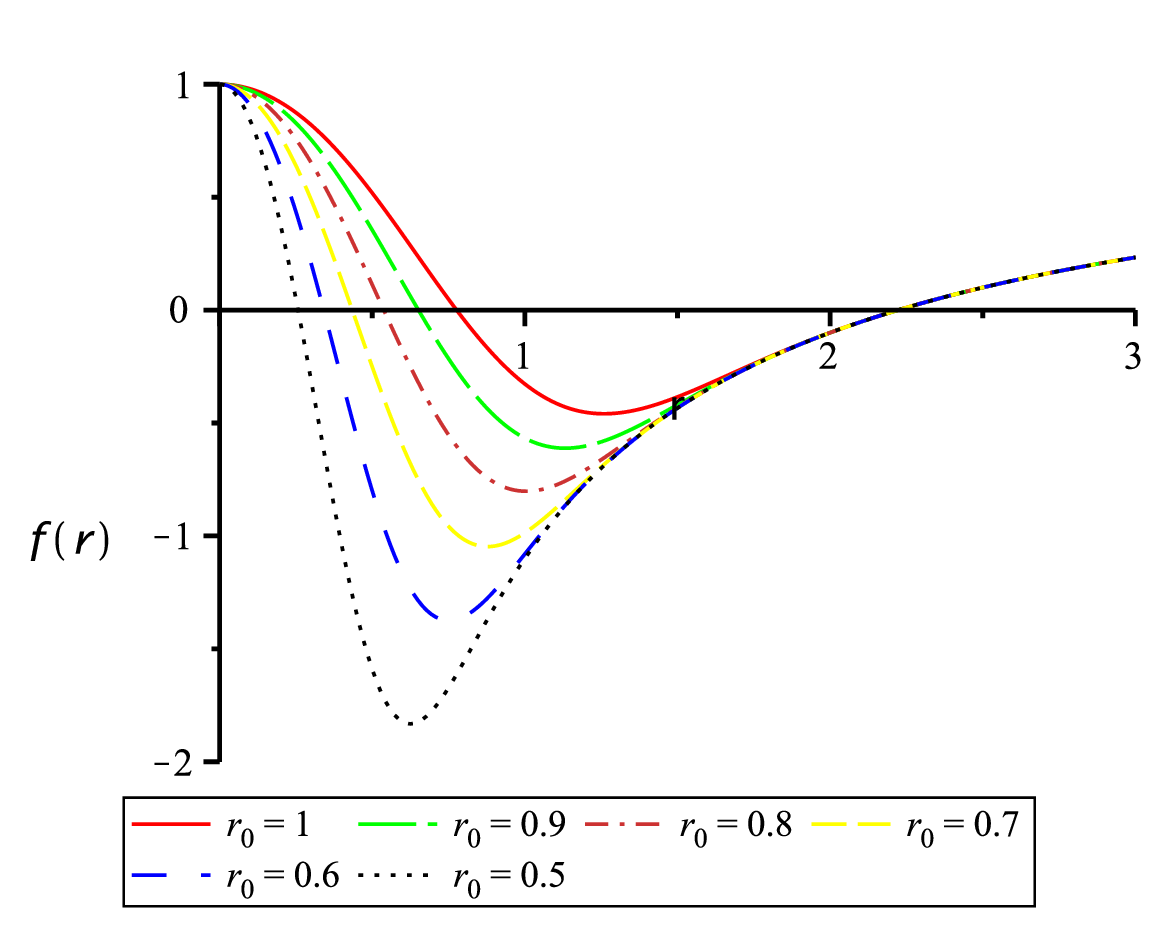}    \includegraphics[width=0.45\textwidth]{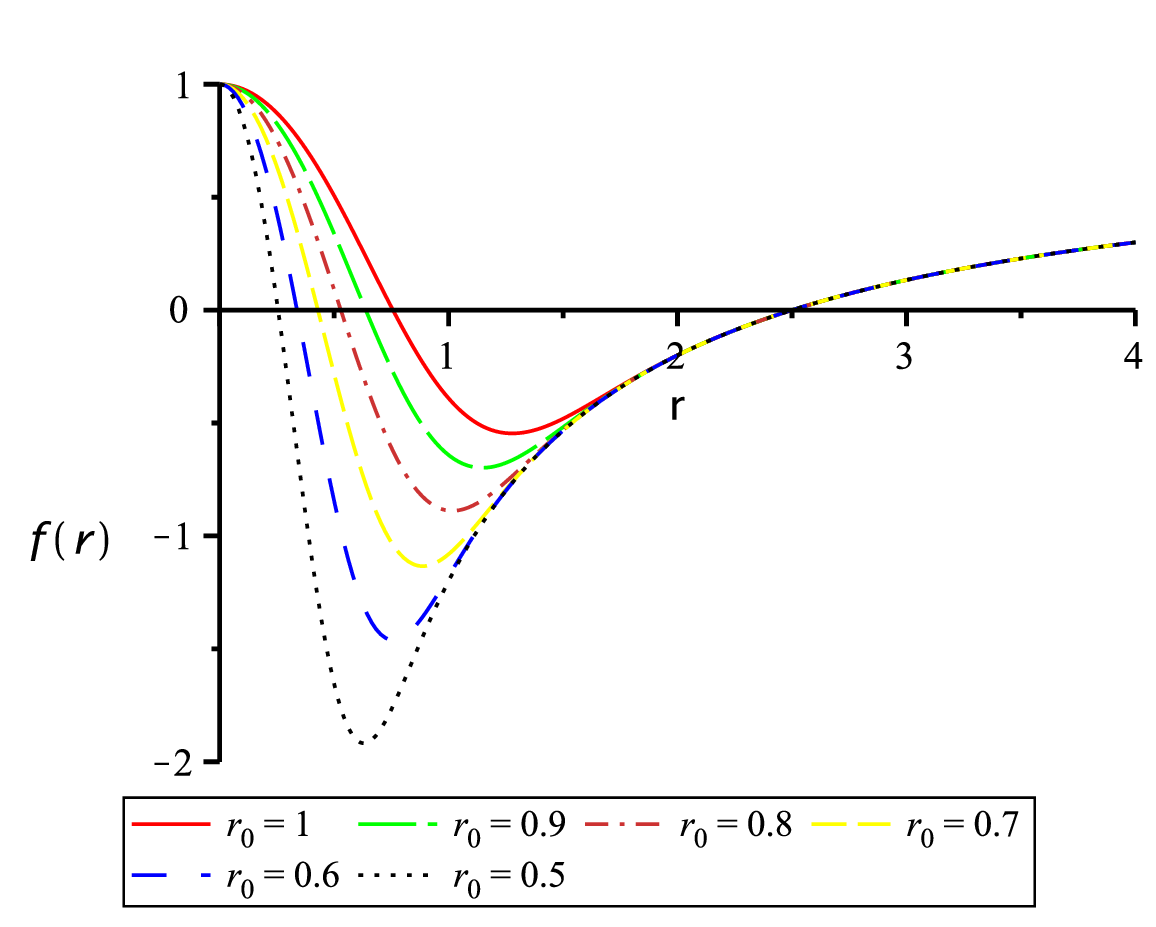}\\
\vspace{1cm}    \includegraphics[width=0.45\textwidth]{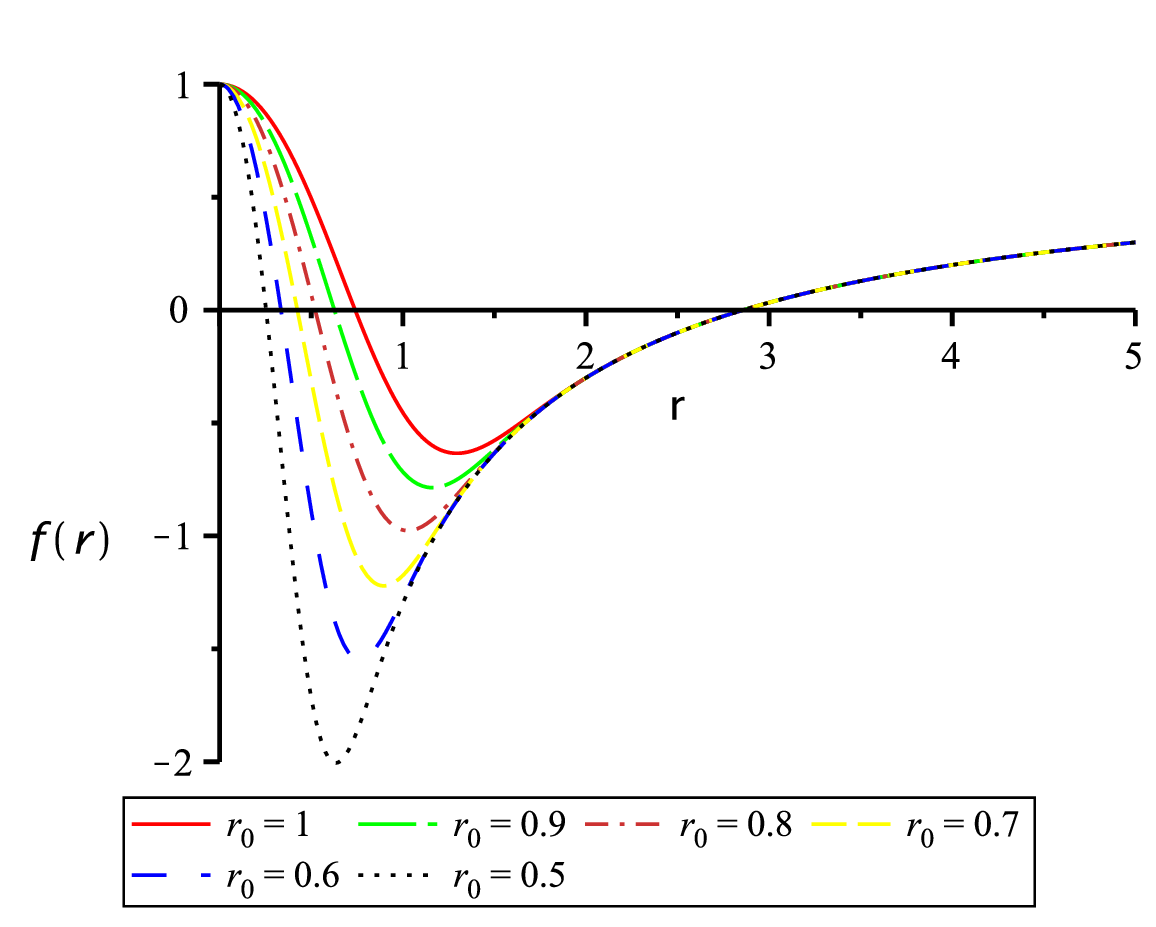}  \includegraphics[width=0.45\textwidth]{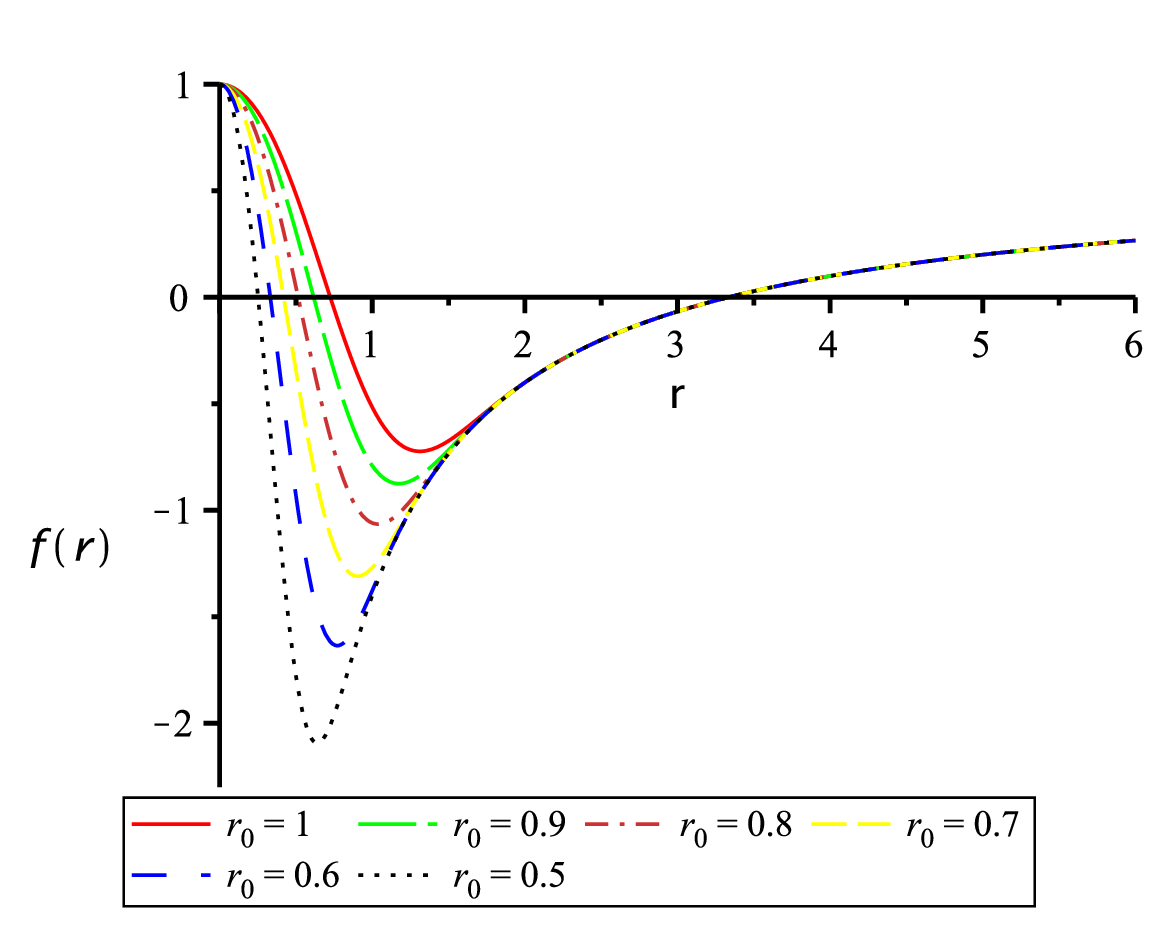}
    \caption{The set of four graphs depicts the behavior of the function \( f(r) \) for different values of the parameter \( \epsilon \). The value of \( \epsilon \) increases from 0.1 to 0.4 in steps of 0.1, starting from the top-left graph to the bottom-right. Specifically, the top-left graph corresponds to \( \epsilon = 0.1 \), the top-right to \( \epsilon = 0.2 \), the bottom-left to \( \epsilon = 0.3 \), and the bottom-right to \( \epsilon = 0.4 \). Each curve within the graphs corresponds to different values of \( r_0 \) (ranging from 1 to 0.5), as indicated in the legend. The points where the curves cross the line \( f(r) = 0 \) represent the horizons of the black hole.}
    \label{f5}
\end{figure*}

In particular, when $r \gg r_0$ this equations reduces to  $\rho(r) \approx \epsilon/8\pi r^2$ corresponding to the energy density for a singular black hole with cloud of strings. In Fig. \ref{f5}, the four graphs illustrate the behavior of the function \( f(r) \) for varying values of the parameter \( \epsilon \). The value of \( \epsilon \) ranges from 0.1 to 0.4. The curves in each graph represent different values of \( r_0 \), varying between 1 and 0.5, as shown in the legend. The intersections of the curves with the line \( f(r) = 0 \) indicate the horizons of the black hole. As we can see, this solution has two horizons, considering the parameter used in each graph. 

\begin{figure*}[t]
    \centering    \includegraphics[width=0.45\textwidth]{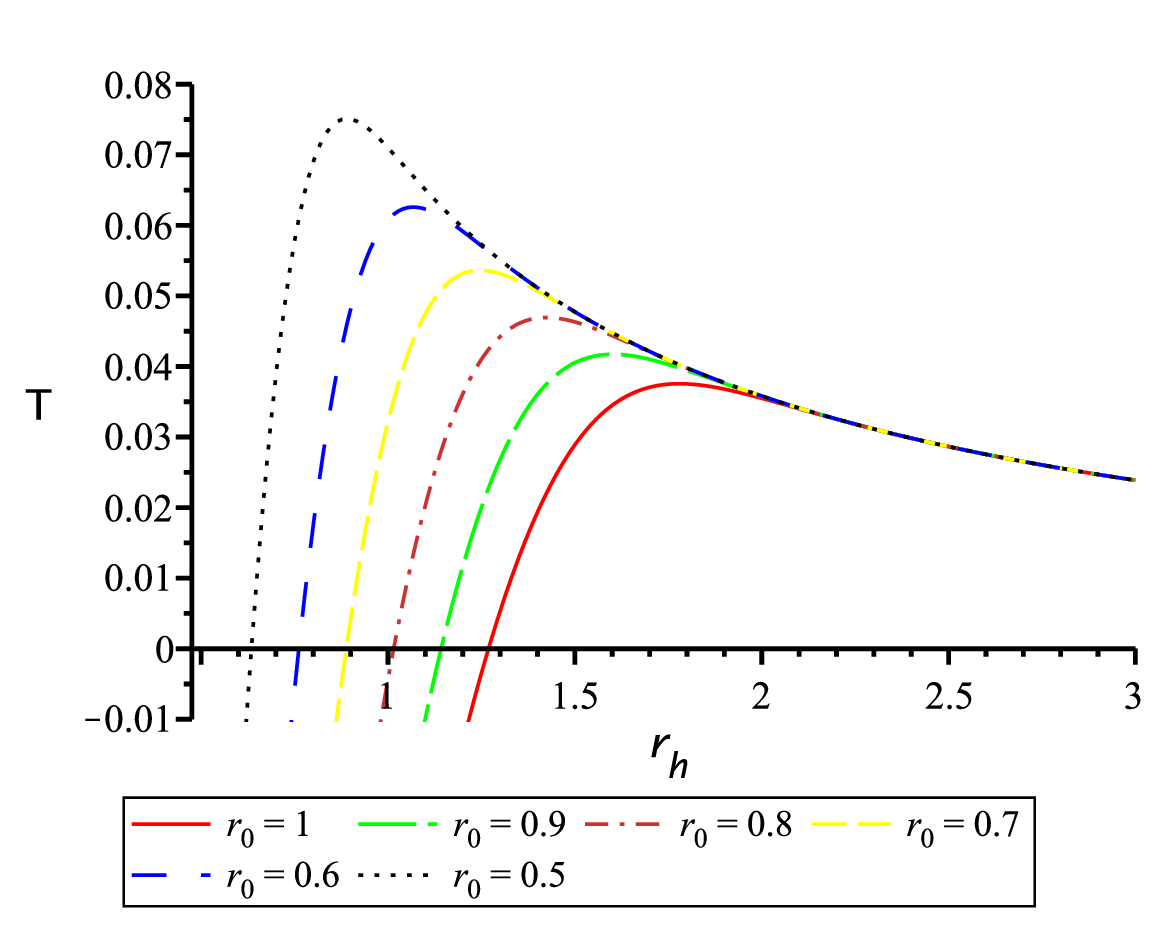}    \includegraphics[width=0.45\textwidth]{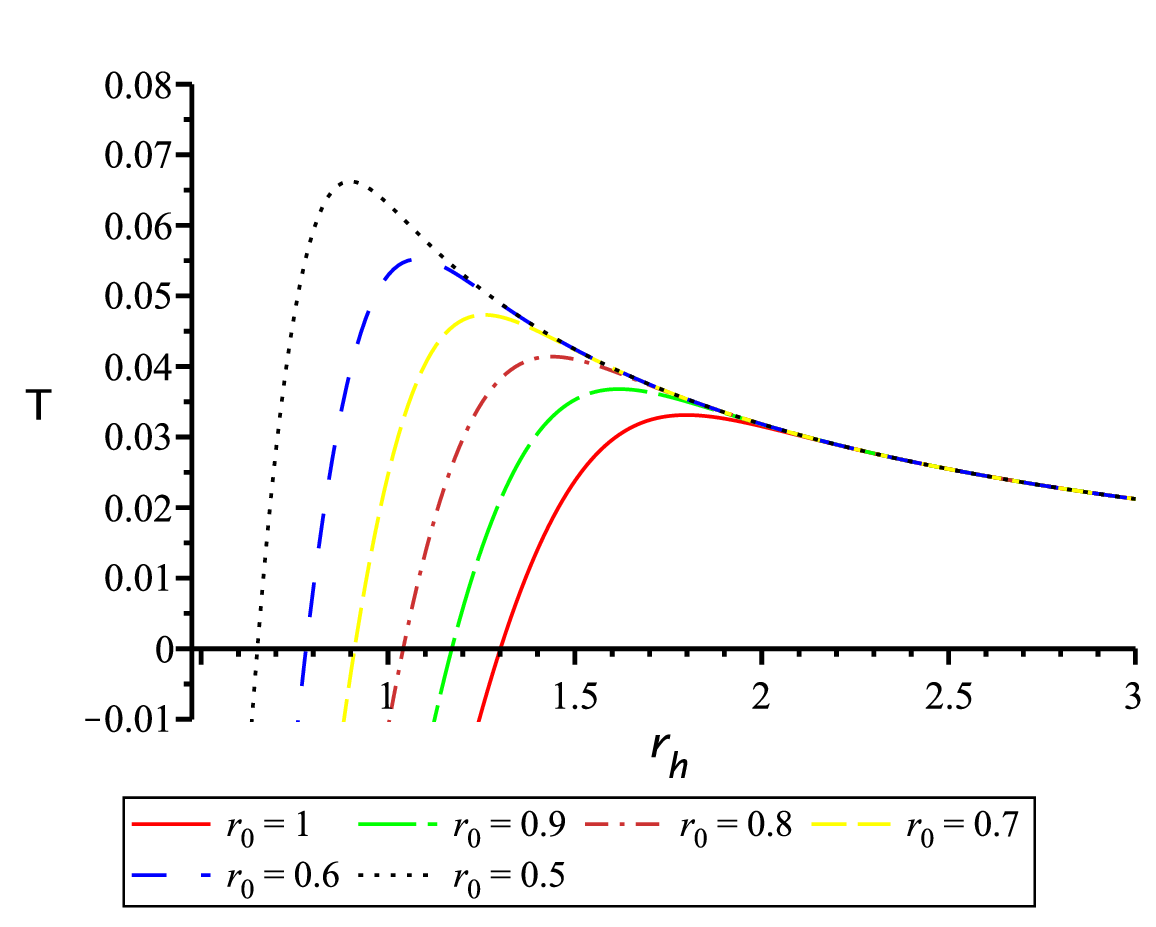}\\
\vspace{1cm}    \includegraphics[width=0.45\textwidth]{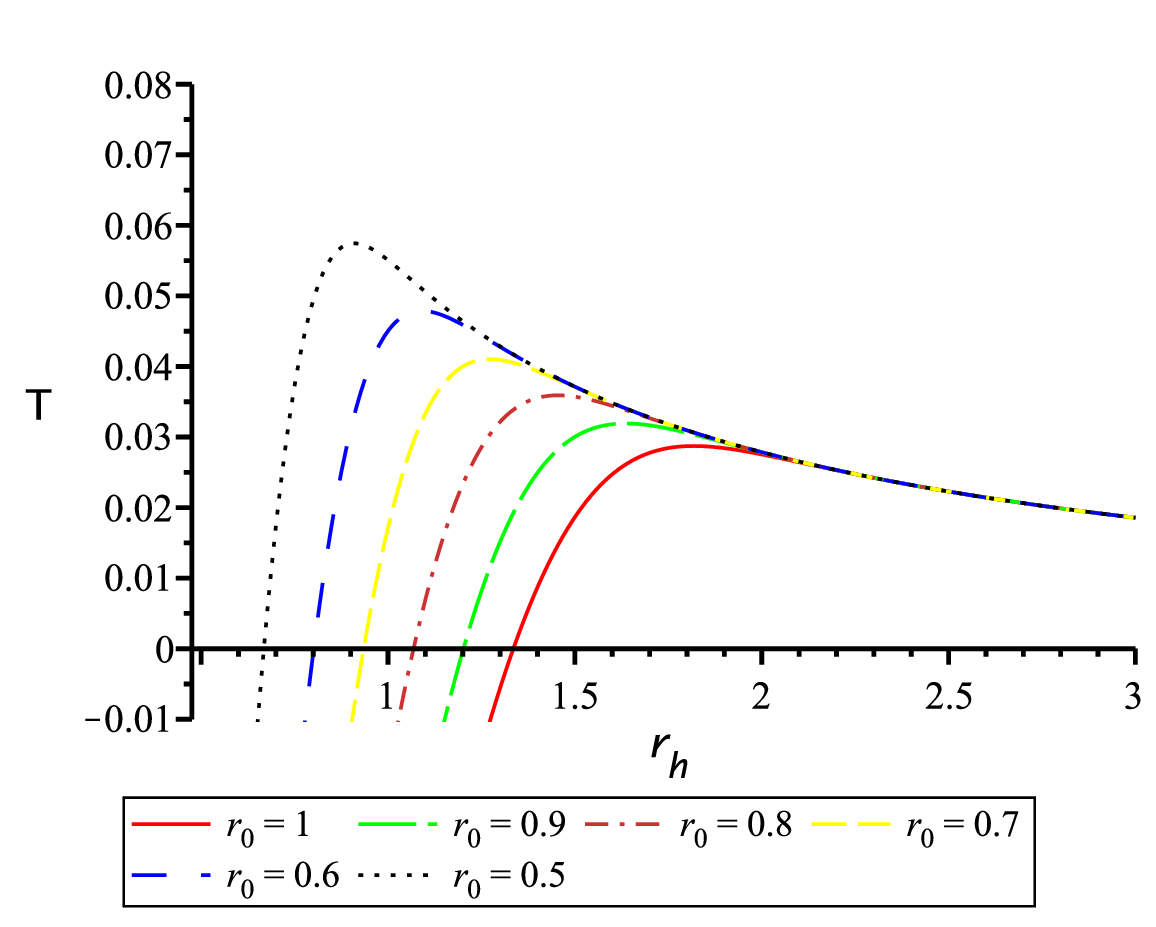}  \includegraphics[width=0.45\textwidth]{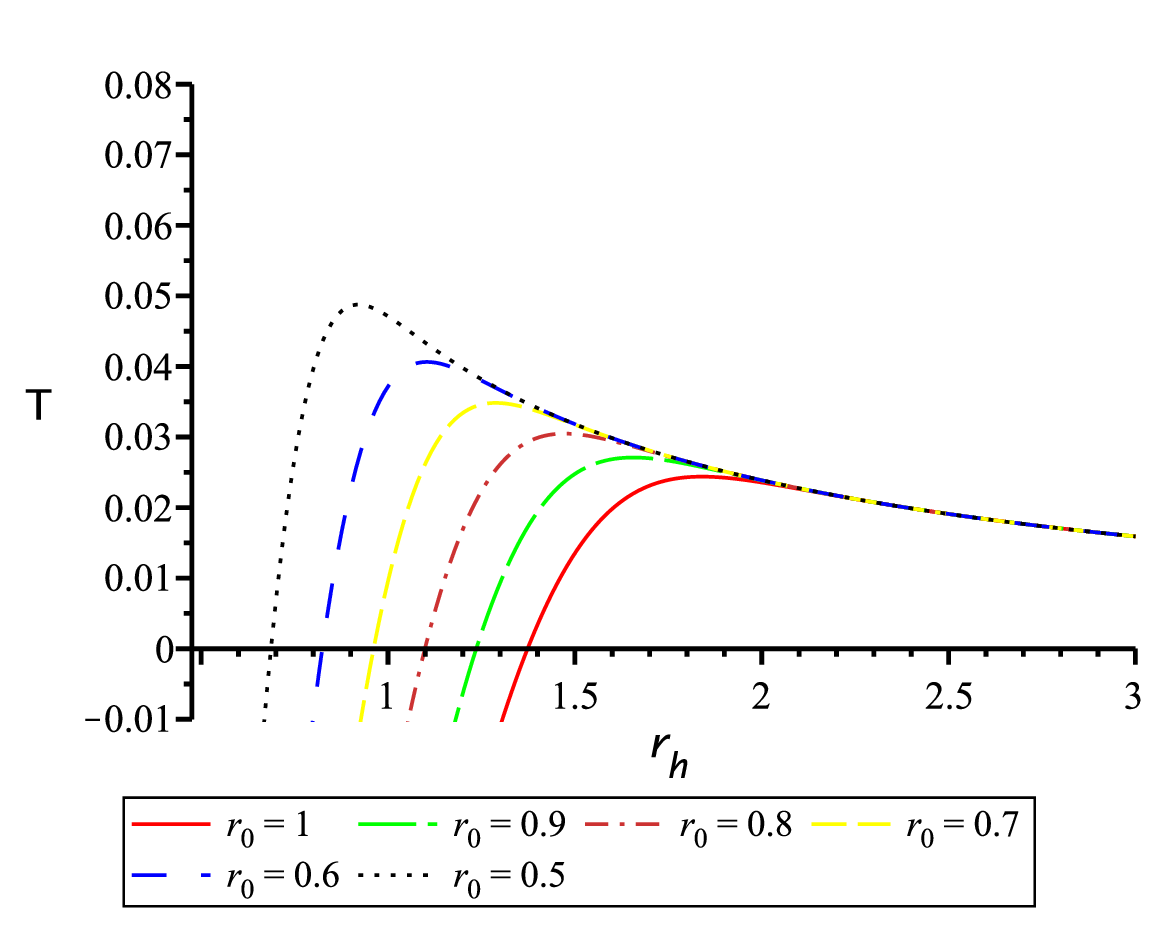}
    \caption{The set of four graphs shows the Hawking temperature as a function of the horizon radius. The value of \( \epsilon \) increases from 0.1 to 0.4 in steps of 0.1, starting from the top-left graph and continuing to the bottom-right. Specifically, the top-left graph corresponds to \( \epsilon = 0.1 \), the top-right to \( \epsilon = 0.2 \), the bottom-left to \( \epsilon = 0.3 \), and the bottom-right to \( \epsilon = 0.4 \). The points where the curves intersect the line \( T = 0 \) indicate negative temperatures, which can arise in black holes with more than one horizon.}
    \label{f6}
\end{figure*}

\begin{figure}[t]
    \centering    
    \includegraphics[scale=0.4]{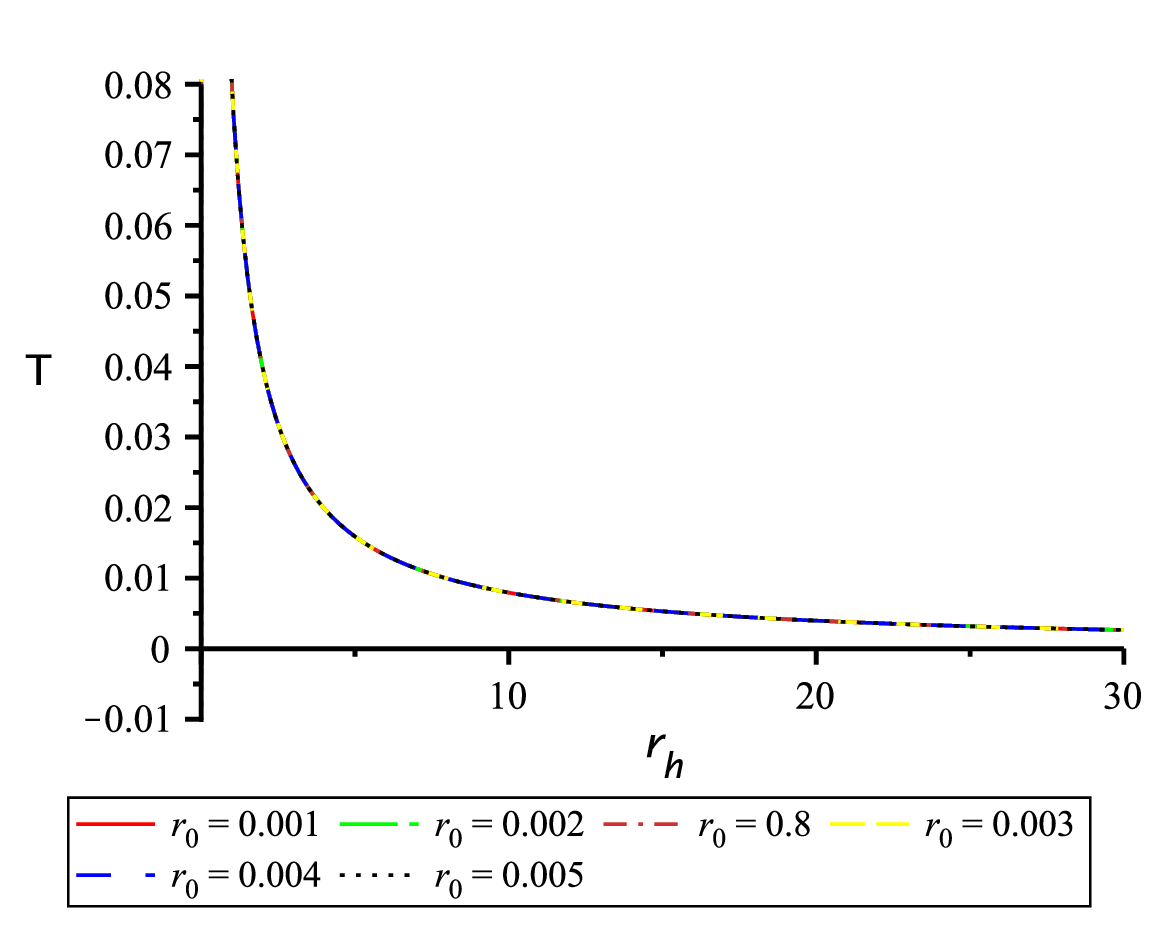}  \caption{This graph displays the Hawking temperature \( T \) as a function of the horizon radius \( r_h \). The points where the curves intersect the line \( T = 0 \) indicate negative temperatures. All the curves exhibit the same shape, demonstrating a rapid decrease in temperature as the horizon radius \( r_h \) increases, eventually asymptotically approaching zero. The initial high temperatures for small \( r_h \) quickly drop, showing that larger black holes have lower temperatures. The uniform behavior across all curves suggests that the parameter variations do not significantly affect the overall shape or trend of the temperature as a function of the horizon radius. This indicates that, despite any differences in parameter \( r_0 \) for small values, the general thermodynamic behavior remains consistent, with temperature decreasing monotonically for increasing horizon radii.}
    \label{f7}
\end{figure}
In Fig. \ref{f6}, four graphs represent the Hawking temperature as a function of the horizon radius obtained by the surface gravity $\kappa=\left.\frac{1}{2}\frac{df(r)}{dr}\right|_{r=r_h}$ of the regular solution. The parameter \( \epsilon \) increases from 0.1 to 0.4 in steps of 0.1, starting with the top-left graph and moving to the bottom-right. In particular, \( \epsilon = 0.1 \) corresponds to the top-left graph, \( \epsilon = 0.2 \) to the top-right, \( \epsilon = 0.3 \) to the bottom-left, and \( \epsilon = 0.4 \) to the bottom-right. The graphs show that the Hawking temperature reaches a peak and then decreases as the horizon radius \( r_h \) increases. For smaller values of \( r_0 \), the maximum temperature is lower, and the peak shifts to larger values of \( r_h \). Additionally, the intersection of the curves with the line \( T = 0 \) indicates a transition to negative temperatures, which can occur in black holes with multiple horizons. As \( \epsilon \) increases, the temperature curves exhibit slight shifts, with the behavior becoming more prominent for larger values of the parameter. This suggests that the parameter \( \epsilon \) influences the thermodynamic behavior of the black hole, possibly altering the location and nature of the horizons.
The case where $r \gg r_0$ is shown in Fig. \ref{f7}. In this case the Hawking temperature has behavior similar to usual singular Schwarzschild black hole. This graph presents the Hawking temperature \( T \) as a function of the horizon radius \( r_h \) for small values of $r_0$. All the curves share a similar shape, showing a decline in temperature as the horizon radius \( r_h \) grows. The initially high temperatures for small \( r_h \) rapidly decrease, indicating that larger black holes tend to have lower temperatures. The consistent behavior across all curves suggests that variations in parameter \( r_0 \) have little impact on the overall trend, reinforcing that the thermodynamic pattern remains unchanged, with temperature steadily decreasing as the horizon radius increases. 

\section{Decomposition of the total energy-momentum tensor for the generalized Fluid of string} \label{sec5}

As observed in \cite{boonserm2016mimicking},  an arbitrary general relativistic anisotropic fluid (spherically symmetric) can be mimicked by an combination of energy-momentum tensors of a perfect fluid $T^{\mu\nu}_f$, an electromagnetic field $T^{\mu\nu}_e$, and a massless minimally coupled scalar field $T^{\mu\nu}_s$ in the form
\begin{align}
\rho &= \rho_f + \frac{1}{2} E^2 + \frac{1}{2} (\nabla \phi)^2, \label{5.1}\\
p_r &= p_f - \frac{1}{2} E^2 + \frac{1}{2} (\nabla \phi)^2, \label{5.2}\\
p_t &= p_f + \frac{1}{2} E^2 - \frac{1}{2} (\nabla \phi)^2,
\label{5.3}
\end{align}
where $\rho_f$ and $p_f$ are the energy density and pressure of fluid respectively, $E(r)$ is the electromagnetic field, and $\phi$ the scalar field. So by interpreting the fluid of strings as an effective anisotropic fluid, we can connect this fluid with tree different types of classical matter. The inverse of Eqs. (\ref{5.1}), (\ref{5.2}) and (\ref{5.3}) can be written as
\begin{align}
p_f &= \frac{1}{2}(p_r + p_t),  \label{5.4}\\
\rho_f &= \rho - \frac{1}{2} \lvert p_r - p_t \rvert,  \label{5.5}\\
(\nabla \phi)^2 &= \max\{p_r - p_t, 0\},  \label{5.6}\\
E^2 &= \max\{p_t - p_r, 0\}.
 \label{5.7}
\end{align}
Thus, we use these results to calculate the values of physical quantities $\phi$ and $E(r)$ associated with the effective anisotropic fluid obtained in this paper. Naturally, these values depend on the form of the energy density and pressures connected by equation of state function $\alpha(r)$. 
\subsection{Reduced fluid of strings}
Let us analyze the fluid of strings solution where $M = 0 $ and $0
 < \alpha < \infty$. By considering the null energy condition (NEC) $\rho + p_i \geq 0$, we get two expressions
\begin{equation}
    \rho + p_r \geq 0\:\: \text{and}\:\:  \rho + p_t \geq 0 \label{5.8}.
\end{equation}
The first equation is trivially satisfied while the second condition is written as
\begin{equation}
    -\epsilon \left(\frac{l}{r}\right)^{\frac{2}{\alpha}} \frac{\alpha + 1}{8 \pi r^2 \alpha} \geq 0,
    \label{5.9}
\end{equation}
as a result, $\epsilon (\alpha+1)/\alpha \leq 0$. If this condition is satisfied, then $p_r - p_t <0$, therefore Eq. (\ref{5.6}) demands  $(\nabla \phi)^2 = 0$ and Eq. (\ref{5.7}) can written in the form
\begin{equation}
   E(r) =  \sqrt{-2\epsilon \left(\frac{l}{r}\right)^{\frac{2}{\alpha}} \frac{\alpha + 1}{16\pi r^2 \alpha}},
   \label{5.10}
\end{equation}
where $\epsilon <0.$ Equation (\ref{5.10}) corresponds to the electric field strength associated with the fluid of strings. In contrast, if NEC is not satisfied, then $p_t - p_r < 0$. In this case, $E = 0$ and 
\begin{equation}
  (\nabla \phi)^2 =  2\epsilon \left(\frac{l}{r}\right)^{\frac{2}{\alpha}} \frac{\alpha + 1}{16\pi r^2 \alpha},
   \label{5.11}  
\end{equation}
where $\epsilon >0$. Since scalar field depends only on $r$, we can write
\begin{equation}
    (\nabla \phi)^2 = g^{rr}(\partial_r\phi)^2 = 2\epsilon \left(\frac{l}{r}\right)^{\frac{2}{\alpha}} \frac{\alpha + 1}{16\pi r^2 \alpha}. 
  \label{5.12}  
\end{equation}
\textit{i.e.,}
\begin{equation}
    \partial_r\phi  = \sqrt{\frac{1}{g_{rr}} 2\epsilon \left(\frac{l}{r}\right)^{\frac{2}{\alpha}} \frac{\alpha + 1}{16\pi r^2 \alpha}}. 
  \label{5.13}
  \end{equation}
In this way, Eq. (\ref{5.13}) can be integrated, the result is
\begin{equation}
     \phi(r)  = \int{\sqrt{\frac{1}{g_{rr}} 2\epsilon \left(\frac{l}{r}\right)^{\frac{2}{\alpha}} \frac{\alpha + 1}{16\pi r^2 \alpha}}}dr.
\end{equation}
Thus, if we consider the NEC, the connection between fluid of strings and the three types of classical matter is different. In the case where NEC is satisfied the scalar part is zero ($(\nabla \phi)^2 =0)$ , in the case where NEC is violated, then the electric field is zero.
\subsection{Regular black hole surrounded by fluid}
\begin{figure}[t]
    \centering    
    \includegraphics[scale=0.37]{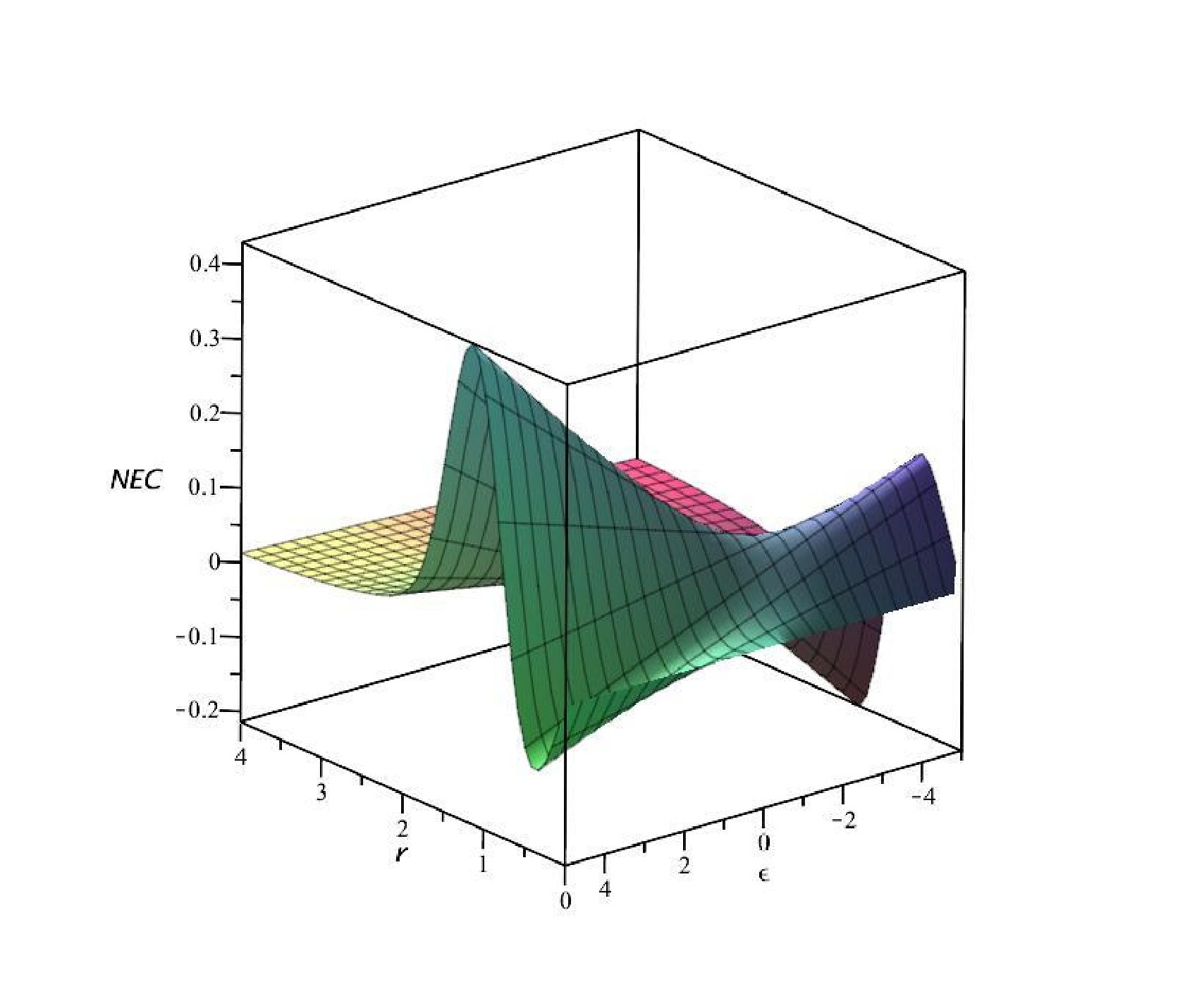}  \caption{The 3D plot shows the variation of the NEC as a function of \(r\) and \(\epsilon\). Positive NEC values indicate regions where the condition is satisfied. In contrast, negative NEC values represent violations of the energy condition. The surface exhibits complex dependencies between the parameters, indicating that values where NEC is satisfied or violation of the NEC depends sensitively on the interplay between \(r\) and \(\epsilon\).}
    \label{f8}
\end{figure}

Considering the solution obtained here associated with a regular black hole surrounded by fluid, the conditions for the NEC provide
\begin{equation}
    \frac{\left( \left( 9 r^6 - 6 r^3 r_0^3 - 2 r_0^6 \right) \epsilon + 9 r^5 r_g \right) \exp\left( -\frac{r^3}{r_0^3} \right) + 2 \epsilon r_0^6}{16 r^2 r_0^6 \pi} \geq 0.
\end{equation}
Based on Fig. \ref{f8}, we can observe specific ranges of parameters \(r\) and \(\epsilon\) where the NEC is positive or negative. Positive NEC regions occur primarily for \(r \approx 1\) to \(r \approx 2\) when \(\epsilon\) is near 0. Additionally, for \(\epsilon \geq 2\), the NEC remains positive around \(r \approx 1/2\) to \(r \approx 2\). There are also small isolated regions where NEC is positive for specific combinations of \(r\) and higher values of \(\epsilon\). Negative NEC regions  appear prominently for \(r > 3\) and negative values of \(\epsilon\), indicating that exotic behavior, such as negative energy densities, may dominate in these ranges. The plot shows a complex interplay between \(r\) and \(\epsilon\), with several oscillations. The analysis can be simplified in the case where $\epsilon = 0$, which indicates the absence of the term associated a cloud-like of strings. In this case, NEC is equivalent to
\begin{equation}
    \frac{9 \, r_g \, \exp\left(-\frac{r^3}{r_0^3}\right) \, r^3}{16 \, \pi \, r_0^6} \geq 0.
\end{equation}
Under these circumstances, the NEC  is naturally satisfied, there are no violations. In addition, $p_r - p_t <0$ implying in  $(\nabla \phi)^2 = 0$ and Eq. (\ref{5.7}) can be expressed as
\begin{equation}
   E(r) =  \sqrt{\frac{9 \, r_g \, \exp\left(-\frac{r^3}{r_0^3}\right) \, r^3}{16 \, \pi \, r_0^6}}.
\end{equation}
Taking the limits $r \gg r_0$ and $r \ll r_0$, it is possible to show that in both limits the result is zero. This suggests that the function reaches a maximum value between the extremes. To find this value we can differentiate $E(r)$ and set it equal to zero, the result of this manipulation is the extreme point of the curve given by $r = r_0$. In conclusion, if $r_0 \rightarrow 0$, the the electric field strength connected to this anisotropic fluid has a maximum value at $r=0$.


\section{Geodesic completeness and shadows of the regular black Hole} \label{geodesic}
Now our focus is on the new solution corresponding to a extension of Dymnikova black hole that recovers the singular black hole surrounded by a cloud of string for large $r$. Here we will analyze the motion of particles in the spacetime generated by this metric. By obtaining constant quantities in the particle trajectories we can derive the equations governing the motion. In particular we can study the motion of massless particles, depending on the choice of parameters of such equations. 
For a black hole to be regular, in addition to the regularity of the curvature invariants, it is necessary that the geodesics in this space-time are complete, that is, the movement of the particles must have no end. This can be evaluated by proper time. For infinite values of proper time the geodesics are complete. Otherwise, it may be necessary to extend to negative values of $r$.
The extension to negative values of r is discussed in addition to the motion of photons in the region associated with the photosphere.
\subsection{Geodesic equations}
\begin{figure}[t]
    \centering    
    \includegraphics[scale=0.44]{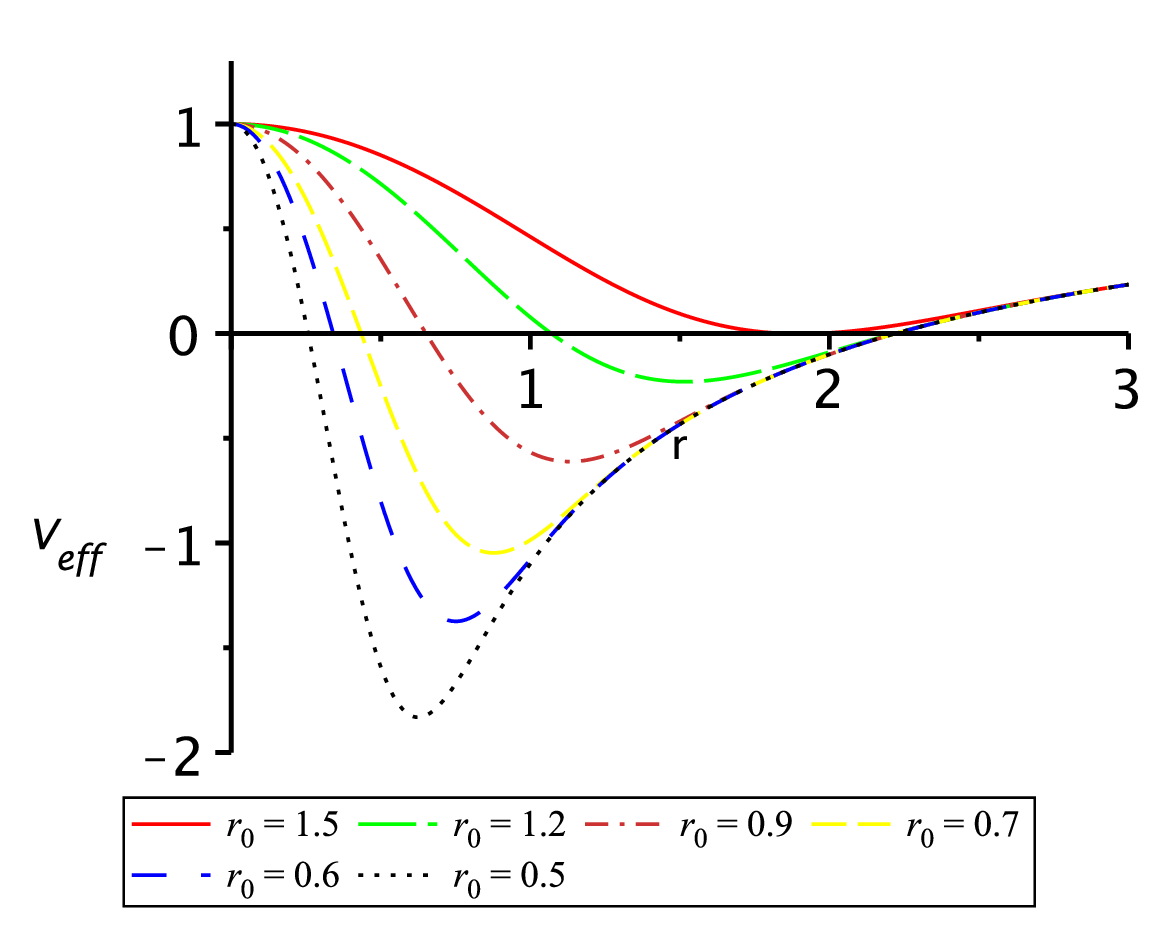}  \caption{The figure illustrates the effective potential \( V_{\text{eff}} \) as a function of the radial coordinate \( r \) for various values of the parameter \( r_0 \). The parameters used are \( L = 0 \), \( E = 2 \), \( \epsilon = 0.1 \), and \( M = 1 \). Each curve corresponds to a specific value of \( r_0 \), as indicated in the legend. The plot highlights the dependency of the effective potential on \( r_0 \), showing distinct behaviors for different values, including turning points and the overall shape of \( V_{\text{eff}} \) in the range of \( r \) from 0 to 3.}
    \label{fig9}
\end{figure}

We begin by expressing the four-velocity as follows:
\begin{equation} u^{\mu}=\frac{dx^{\mu}}{d\tau}=\Dot{x}^{\mu}. \label{3c1} \end{equation}
Due to spherical symmetry, we can consider the motion executed in the equatorial plane by making $\theta =\text{constant} =\pi/2$ so that the analysis of the motion of null geodesics is not affected. From equation (\ref{3c1})  equation, the scalar product can be written as
\begin{equation} u^{\mu}u_{\mu}=-\kappa=-f(r)\Dot{t}^2+f(r)^{-1}\Dot{r}^2 + r^2\Dot{\varphi}^2. \label{3c2} \end{equation}
Here, $\kappa=1$ represents time-like geodesics, while $\kappa=0$ corresponds to null geodesics. The spacetime characterized by the metric function (\ref{302}) possesses two Killing vectors, specifically $t^{\mu}=(1,0,0,0)$ and $\varphi^{\mu}=(0,0,0,1)$. These Killing vectors are associated with two conserved quantities, namely the energy $E$ and the angular momentum $L$. These quantities are related through the equations:
\begin{equation} -g_{\mu\nu}t^{\mu}u^{\nu}=f(r)\Dot{t}=E,\:\:\: g_{\mu\nu}\varphi^{\mu}u^{\nu}=r^2\Dot{\varphi}=L. \label{3c3} \end{equation}
By substituting the above relations into Equation (\ref{3c2}), we derive:
\begin{equation} -\kappa=-\frac{E^2}{f(r)}+\frac{\Dot{r}^2}{f(r)} + \frac{L^2}{r^2}, \label{3c4} \end{equation}
which can be rearranged into the following form
\begin{equation} \Dot{r}^2 + V(r)= E^2, \label{3c5} \end{equation}
where the effective potential $V(r)$ is defined as:
\begin{equation} V(r)=\frac{L^2 f(r)}{r^2}+\kappa f(r). \label{3c6} \end{equation}
In Fig. \ref{fig9}, it is shown the shape of effective potential for solution (\ref{302}) for $\kappa=1$.  As we can see, this space-time has two horizons. However, at critical $r_{0}=1.5$ value the geometry has no horizon. When the coordinate reaches $r=0$, the potential reaches $V_{eff}=1$. The calculation of the invariants at this point provides a regular value. At the core of this spacetime, the metric function can be written as $f(r) \approx 1-(r_g/r_0^3)r^2$, what is expected for the matter in this region. Due to the high density it is expected that quantum effects take place in the geometry. As a result, the  geometry is described by a effective regular de Sitter core \cite{dymnikova2020dark}. 
\begin{figure}[t]
    \centering    
    \includegraphics[scale=0.44]{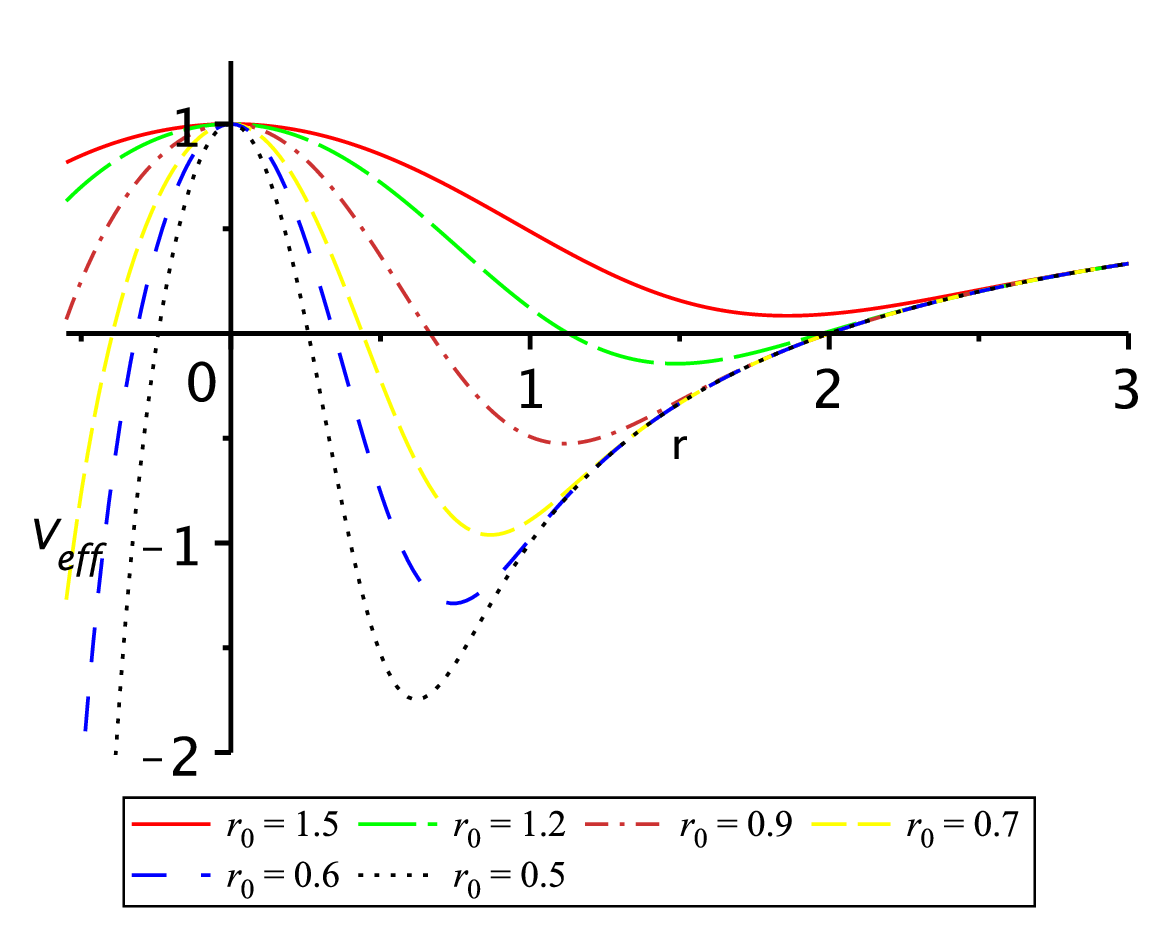}  \caption{The figure depicts the effective potential \( V_{\text{eff}} \) as a function of the radial coordinate \( r \), where \( r \) can take both positive and negative values. The parameters used are \( L = 0 \), \( E = 2 \), \( \epsilon = 0 \), and \( M = 1 \). Each curve represents a specific value of \( r_0 \), as indicated in the legend (\( r_0 = 1.5, 1.2, 0.9, 0.7, 0.6, 0.5 \)). 
The plot demonstrates the variation of \( V_{\text{eff}} \) with \( r \), showing the behavior for positive and negative \( r \) near $r=0$. It also illustrates how the shape and turning points of the potential depend on the chosen values of \( r_0 \), particularly emphasizing the influence of \( r_0 \) on the potential's depth.}
    \label{fig10}
\end{figure}
\subsection{Geodesic completeness}
For massive particles and purely radial motion, we choose a path where the equation for $r$ becomes
\begin{equation}
    \dot{r}^2 = E^2 - f(r),
\end{equation}
and by integrating this expression, we obtain the proper time
\begin{equation}
    \tau = \int_{r}^{r_i}dr\frac{1}{\sqrt{E^2 - f(r)}},
    \label{tau}
\end{equation}
where $r_i$ is the initial radial position. For $E=1$. integral (\ref{tau}) can be solved as $\tau \propto  \ln(r)$, this means that a particle needs a infinity proper time to arrive at $r=0$, a similar result can be obtained for $E>1$. Thus, the proper time assumes infinite value in the range  $r \in [0,\infty)$ as a result. This indicates that the motion of the particle does not end, \textit{i.e.,} the geodesics are complete considering this range. Note that if the energy is $E < 1 $, the particle cannot cross the potential
barrier and will bounce back. A detailed analysis for other regular black holes can be found in \cite{modesto,modesto2}. Note that in the case of the extension obtained here where $\epsilon \neq 0$, the asymptotic limits are equivalent to the Dymnikova black hole, consequently the discussed results are similar for the  Dymnikova geometry. 
\subsection{Extension to negative
values of the radial coordinate}
The extension to negative values may be required for a complete analysis of geodesic motion in the geometry \cite{modesto}. For the case of the regular black hole obtained here, the infinite value of the proper time obtained to reach $r=0$, does not imply the need for extension to negative values. In fact, considering the Dymnikova black hole for simplicity ($\epsilon =0$). In the $r \rightarrow 0^{+}$ limit, our function reads $f(r)=1$. Similarly, in the $r \rightarrow 0^{-}$ limit, the function is $f(r)=1$. Thus the metric function is continuous at $r=0$. In Fig. \ref{fig10} the plot shows the extension of effective potential for negative values of $r$. As we can see, the potential is not symmetric around $r=0$. Calculating the proper time from a starting point to $r \rightarrow -\infty$ gives a finite value, indicating that the geodesics are not complete in this region. 
\subsection{Null geodesics and shadows of regular Black Hole}
\begin{figure}[t]
    \centering    
    \includegraphics[width=.5\textwidth]
    {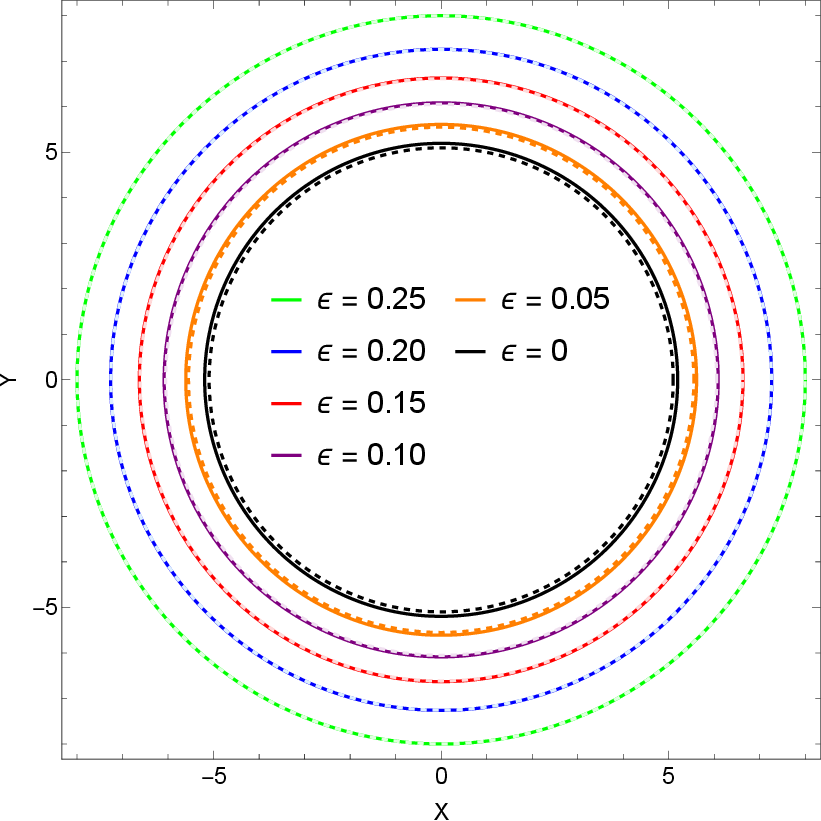}  \caption{This graph represents the shadows of the regular black hole solution ($\ref{302}$). Dashed lines correspond to the case where \( r_0 = 1.9 \), while solid lines correspond to \( r_0 = 1.1 \). Lines of the same color represent the same value of \( \epsilon \). The black color represents the Dymnikova solution \cite{dymnikova1992vacuum}.}
    \label{f20}
\end{figure}
The last equation in (\ref{3c3}) can be combined with equation (\ref{3c5}) to give the null particle orbits that depend on $r$ and $\varphi$
\begin{equation}
    \left(\frac{dr}{d\varphi}\right) = \pm r \sqrt{\frac{r^2}{b^2} - f(r)},
    \label{d}
\end{equation}
where $b \equiv L/E$ is the impact parameter for null particles. The closest approach $\bar{r}_0$ can be defined by the point where $dr/d\varphi = 0$. This expression used in Eq. (\ref{d}) gives
\begin{equation}
    b = \frac{\bar{r}_0}{\sqrt{f(\bar{r}_0)}}. 
    \label{dd}
\end{equation}
To calculate the Shadows of black holes, we will need the critical impact parameter, defined as
\begin{equation}
    b_c(r_m) \equiv \lim_{\bar{r}_0\rightarrow r_m}{b(\bar{r}_0)}, 
    \label{ddd}
\end{equation}
where $r_m$ corresponds to the extremum values of the curves of the effective potential and is the lower bound for stable orbits. In the case of the Schwarzschild geometry $r_m = 3M$, corresponding the black hole photon sphere. For the regular black hole given by function ($\ref{302}$), $r_m$ is obtained from the extremum of the effective potential $dV_{\text{eff}}/dr=0$, such that its value is approximately given by $r_m \approx 3r_g/(2-2\epsilon)$, in the case of $\epsilon = 0$ the value of $r_m$ coincides with the Schwarzschild photon sphere, as expected. The shadows associated with our regular black hole, are obtained by $b_c$ of geometry ($\ref{302}$). The result is 
\begin{equation}
  b_c = \frac{9 \, r_g}{2( 1-\epsilon )\sqrt{3 \, \, (2 + \epsilon) \, e^{\frac{27 \, r_g^3}{8 \, (-1 + \epsilon)^3 \, r_0^3}} + 3(1-\epsilon)}}.
\end{equation}
In this way, in the case of $r \rightarrow \infty$ the angular radius of the shadows $\alpha$, defined as the opening angle of the cone shadow \cite{perlick2022calculating} can written as
\begin{equation}
    \alpha = \frac{b_c}{r_{D}},
\end{equation}
where $r_{D}$ is the observer distance. Fig. \ref{f20} depicts the critical impact parameter radius of the regular black hole in Cartesian coordinates. The parameter \(r_0\) influences the size of the shadows, with dashed lines corresponding to \(r_0 = 1.9\) and solid lines to \(r_0 = 1.1\). As we can see, larger values of $\epsilon$ produce larger shadows sizes.  
Data obtained by the Event Horizon Telescope (EHT) collaboration provide a value for the total angular diameter $\theta_d = 2\alpha$ of $48.7 \pm 7\mu as$ and $42 \pm 3\mu as$ for the black holes M87* and Sgr A* respectively \cite{eht1,eht2}. For M87* black hole, we can reproduce angular diameter with solution ($\ref{302}$) using the values $r_0=1.54\times 10^{10}\: m$ and $\epsilon = 0.035$. In the Sgr A* case, the best values are $r_0=1.54\times 10^{10}\: m$ and $\epsilon = 0.05$.  

\section{concluding remarks} \label{sec6}

In Section \ref{sec2}, we studied the cloud of strings model initially proposed by Letelier, which was later generalized to incorporate pressure. We replaced the usual constant parameter $\alpha$ with a function that varies with the radial coordinate, allowing for a broader set of solutions, including those representing energy densities associated with both standard and exotic matter. 

Through this approach, we obtained two main classes of solutions: one where $\alpha$ remains constant and another where it varies with $r$. For the case $\alpha = 2$, a logarithmic solution was identified, which may have astrophysical relevance as it produces flat rotation curves at large radii \cite{soleng1995dark}. This class of solutions demonstrates that the interplay between gravity and fluid of strings can yield novel configurations while also recovering Schwarzschild-like solutions in the absence of a fluid. The importance of $\alpha$ as a controlling parameter for the structure of the space-time was highlighted, especially in determining horizon locations.

In Section \ref{sec3}, We obtain a particular solution considering $\alpha =$ constant inspired by an equivalent solution in the context of Kiselev black holes. We showed that Kiselev solution, which describes a black hole surrounded by an anisotropic effective fluid, shares structural similarities with the fluid of strings. By comparing their energy-momentum tensors, we found that the two models are mapped through the relation $\alpha \rightarrow 2 / (3w + 1)$, where $w$ is the parameter of the equation of state for Kiselev fluid.
The physical implications of this connection were explored by examining how the radial and tangential pressures change in these systems. We found that the tangential pressure in the strings fluid changes sign in different regions compared to Kiselev fluid. Afterwards, we derived a new class of regular black hole solutions by coupling the fluid of strings with an exponential term in the metric function. This modification ensured that the resulting black holes remain regular at the origin, unlike the usual singular Schwarzschild solution with cloud/fluid of strings. The new solution reduces to the Schwarzschild black hole with a cloud of strings in the limit of large radial distances. The exponential coupling with the fluid parameter $\epsilon$ allows new physical results associated with the horizon and thermodynamic properties. 

The thermodynamic properties of these solutions were studied by calculating the Hawking temperature as a function of the horizon radius and fluid parameters. We found that the temperature exhibits non-trivial behavior, with its peak shifting based on the value of $\epsilon$. Larger values of $\epsilon$ correspond to lower peak temperatures and more gradual increases in temperature with the horizon radius. This suggests that the presence of the fluid of strings with variable equation of state influences the thermodynamics of the black hole, making it significantly different from the standard Schwarzschild black hole. In systems with multiple horizons, the temperature can even become negative, indicating the presence of more complex physical phenomena. The results show that larger black holes tend to have lower temperatures, consistent with the general behavior of black holes. Finally, we have expressed the energy-momentum tensor for the fluid of strings as a combination of three types of energy-momentum tensors: a perfect fluid, an electromagnetic field, and a scalar field. 
Our findings are summarized below:

\begin{itemize}
    \item We propose a generalization of equation of state used in \cite{soleng1995dark}. We consider that $\alpha$ depends on the radial coordinate, \textit{i.e.}, $\alpha = \alpha(r)$. 
    \item For $\alpha = 2$, a logarithmic solution emerges, which is relevant for modeling galactic rotation curves due to the flat behavior at large radii.  
    \item Based on the connection with Kiselev anisotropic fluid, we propose a solution of fluid of strings associated with the range $0 < \alpha < \infty$ and with $2M = 0$.  We referred it as reduced fluid of strings solution.  
    \item A new regular black hole solution was derived by coupling the cloud of strings with an exponential term, ensuring regularity at the origin and recovering Schwarzschild with cloud of strings behavior at large distances.  
    \item We show that Hawking temperature is significantly affected by the fluid parameters, with larger $\epsilon$ values producing lower temperatures and more gradual variations with the horizon radius.  
    \item In systems with multiple horizons, negative temperatures can occur.  
    \item The energy-momentum tensor for the fluid of strings was decomposed into contributions from three components: a  isotropic perfect fluid, an electromagnetic field, and a scalar field minimally coupled 
    \item  We study geodesic motion of particles, geodesic completeness and shadows in the geometry of the novel regular black hole.
\end{itemize}

In future work, we can explore additional aspects of new solutions that describe regular black holes using this approach. Another possibility is to extend the ideas discussed here to the framework of modified gravity theories.
\acknowledgments
L.C.N.S. would like to thank FAPESC for financial support under grant 735/2024. 
\bibliographystyle{apsrev4-1}
\bibliography{referencias_unificadas}

\end{document}